\documentclass[12pt]{article}
\usepackage{scicite}

\usepackage{times}
\usepackage{upgreek}
\usepackage[usenames, dvipsnames]{color}
\usepackage{graphicx}
\usepackage{gensymb}
\usepackage{subfiles}
\usepackage{rotating}
\usepackage{caption}
\usepackage{soul} 
\usepackage{hyperref}
\usepackage{listings}
\usepackage{upgreek}
\usepackage{footnote}

\newcommand{\beginsupplement}{%
        \setcounter{table}{0}
        \renewcommand{\thetable}{S\arabic{table}}%
        \setcounter{figure}{0}
        \renewcommand{\thefigure}{S\arabic{figure}}%
}

\def\apj{Astrophys.~J.}                 
\def\apjl{Astrophys.~J.}                
\def\apjs{Astrophys.~J.~Suppl.~Ser.}               
\def\aap{Astron.~Astrophys}                
\def\mnras{Mon. Not. R. Astron. Soc.}             
\def\prd{Phys.~Rev.~D}        
\def\pasp{PASP}               
\def\pasa{Astron. Soc. Aust.}               
\def\nat{Nature}              

\def\physrep{Phys.~Rep.}

\definecolor{meatbrown}{rgb}{0.9, 0.72, 0.23}
\definecolor{palatinatepurple}{rgb}{0.41, 0.16, 0.38}
\definecolor{outerspace}{rgb}{0.5, 0.5, 0.5}
\definecolor{pear}{rgb}{0.82, 0.89, 0.19}
\definecolor{phthalogreen}{rgb}{0.07, 0.21, 0.14}
\definecolor{aqua}{rgb}{0.0, 1.0, 1.0}
\definecolor{aureolin}{rgb}{0.99, 0.93, 0.0}
\definecolor{babyblue}{rgb}{0.54, 0.81, 0.94}
\definecolor{brilliantrose}{rgb}{1.0, 0.33, 0.64}
\definecolor{sepia}{rgb}{0.44, 0.26, 0.08}
\definecolor{sunglow}{rgb}{1.0, 0.8, 0.2}
\definecolor{myred}{rgb}{0.9, 0.0, 0.0}
\definecolor{pinegreen}{rgb}{0.0, 0.5, 0.0}

\newenvironment{sciabstract}{%
\begin{quote} \bf}
{\end{quote}}

\title{Lense-Thirring frame dragging induced by a fast-rotating white dwarf in a binary pulsar system} 
\author
{
V. Venkatraman Krishnan$^{1,2\dagger}$,  
M. Bailes$^{1,3}$, 
W. van Straten$^{4}$,
N. Wex $^{2}$,\\
P. C. C. Freire$^{2}$, 
E. F. Keane $^{1,5}$,
T. M. Tauris $^{6,7,2}$,
P. A. Rosado$^{1}$\footnote{Present address: Holaluz-Clidom, S.A., Passeig de Joan de Borbó, 99-101, 4a, 08039 Barcelona (Spain)} ,\\
N.D.R. Bhat $^{8}$,
C. Flynn $^{1}$, 
A. Jameson $^{1}$,
S. Os{\l}owski $^{1}$\\
\normalsize{$^{1}$ Centre for Astrophysics and Supercomputing, Swinburne University of Technology}\\
\normalsize{Melbourne, VIC 3122, Australia.}\\
\normalsize{$^{2}$ Max-Planck-Institut f{\"u}r Radioastronomie,  D-53121 Bonn, Germany. }\\
\normalsize{$^{3}$ Australian Research Council Centre of Excellence for Gravitational Wave Discovery (OzGrav), }\\
\normalsize{Swinburne University of Technology, Melbourne, VIC 3122, Australia.}\\
\normalsize{$^{4}$ Institute for Radio Astronomy and Space Research, Auckland University of Technology,}\\ \normalsize{ Auckland 1142, New Zealand.}\\ 
\normalsize{$^{5}$ Square Kilometer Array Organisation, Jodrell Bank Observatory, Macclesfield, SK11 9DL, UK. } \\
\normalsize{$^{6}$ Aarhus Institute of Advanced Studies, Aarhus University, 8000~Aarhus~C, Denmark.}\\
\normalsize{$^{7}$ Department of Physics and Astronomy, Aarhus University, 8000~Aarhus~C, Denmark.}\\
\normalsize{$^{8}$ International Centre for Radio Astronomy Research, Curtin University, Bentley, WA 6102, Australia.}\\ 
\normalsize{$^\dagger$To whom correspondence should be addressed: }\\
\normalsize{E-mail:vkrishnan@mpifr-bonn.mpg.de}
}
 
\date{}

\begin{document} 


\baselineskip24pt

\maketitle 
\noindent

\clearpage
\begin{sciabstract} 
Radio pulsars in short-period eccentric binary orbits can be used to study both gravitational dynamics and binary evolution. The binary system containing PSR~J1141$-$6545 includes a massive white dwarf (WD) companion that formed before the gravitationally bound young radio pulsar. We observe a temporal evolution of the orbital inclination of this pulsar that we infer is
caused by a combination of a Newtonian quadrupole moment and Lense-Thirring precession of the orbit resulting from rapid rotation of the WD. Lense-Thirring precession, an effect of relativistic frame-dragging, is a prediction of general relativity. This detection is consistent with the evolutionary scenario in which the WD accreted matter from the pulsar progenitor, spinning up the WD to a period of less than 200 seconds.
\end{sciabstract}

In general relativity (GR), the mass-energy current of a rotating body induces a gravitomagnetic field, so-called because it has formal similarities with the magnetic field generated by an electric current \cite{CiufoliniWheeler1995}. This gravitomagnetic interaction drags inertial frames in the vicinity of a rotating mass. The strength of this drag is proportional to the body's intrinsic angular momentum (spin). Frame-dragging in binary systems cause precession of the orbital plane, called Lense-Thirring (LT) precession \cite{LenseThirring1918}. The effect has been detected in the weak-field regime of the Earth by satellite experiments in the gravitational field of the rotating Earth \cite{CiufoliniPavlis2004,EverittEtAl2011}. Frame-dragging is also a plausible interpretation for X-ray spectra of accreting black holes, because it affects photon propagation and the properties of the accretion disk, which in some cases, allows the determination of the black hole spin \cite{Reynolds2019}.

In binary pulsar systems [systems containing a rotating magnetized neutron star (NS), the radio emission of which is visible from Earth as a pulsar and an orbiting companion star], relativistic frame-dragging caused by the spin of either the pulsar or its companion is expected to contribute to spin-orbit coupling. These relativistic effects are seen in addition to Newtonian contributions from a mass-quadrupole moment (QPM) induced by the rotating mass\cite{Wex1995}. Both contributions cause a precession of the position of the periastron ($\omega$; the point in the pulsar orbit that is closest to its companion), and a precession of the orbital plane, changing the orbital inclination ($i$; see Figure~1). If these precessional effects are induced by the NS rotation, they are dominated by LT, whereas in the case of a rotating main-sequence companion star, they are dominated by QPM interactions \cite{Wex1995,KaspiEtAl1996}. Fast rotating white dwarf (WD) companions with spin periods of a few minutes fall between these two extremes and are expected to have similar contributions from both effects \cite{SOM}. Although QPM spin-orbit interaction has already been observed in some binary pulsars [for instance, PSR~J0045$-$7319 \cite{KaspiEtAl1996}], no binary pulsar orbit has been shown to experience a measurable contribution from LT drag.

The times of arrival (TOAs) of the radio pulses from pulsars can be measured with high precision, with uncertainties that are often
more than three orders of magnitude smaller than their spin periods. This allows pulsars to be monitored for decades without losing rotational phase information. This ``pulsar timing" methodology can provide precise measurements of their spin and astrometric parameters  \cite{Lorimer&Kramer2005}. For pulsars in binary systems, pulsar timing also provides precise measurements of their orbit: five parameters describing the non-relativistic (Keplerian) parameters and, for some binaries, relativistic effects that affect both the orbit and the propagation of the radio signals \cite{DamourTaylor1992}. These relativistic effects are typically parameterized using theory-independent post-Keplerian (PK) parameters \cite{DD1,DD2}. Measurements of two PK parameters can be used to obtain the mass of the pulsar ($M_{\rm p}$) and of the companion ($M_{\rm c}$) by assuming a theory of gravity such as GR whereas three or more PK parameters can be used to perform self-consistency tests of that theory. An alternative formalism assumes a theory of gravity such as GR, which allows direct model fitting of the component masses. This is preferred if the goal is to understand the properties and dynamics of the system under that theory, rather than testing the theory itself. We adopt this approach in this report, and assume  that GR adequately describes the system.

PSR~J1141$-$6545 is a radio pulsar with a spin period of $\sim394$\, ms in a $\sim$4.74-hour eccentric orbit with a massive WD companion \cite{KaspiEtAl2000,AntoniadisBassaWexEtAl2011}. It is one of only two confirmed NS-WD binary systems (the other being PSR~B2303+46, a wider-orbit binary) in which the WD is known to have formed first and is thus older than the NS. This requires an unusual evolution of the stellar pair \cite{TaurisEtAl2000,ChurchEtAl2006}. The initially more massive (primary) star must have formed the older massive WD. Forming a NS requires a higher mass, so the initially (slightly) less massive secondary star must have accreted sufficient mass from the primary star to explode in a supernova (SN), producing the pulsar. Before exploding, the secondary would have undergone an expansion leading to mass transfer back to the primary star, by that point already a WD.

Since the primary was already a WD, there cannot have been subsequent mass accretion onto the newly formed pulsar. Thus, unlike most  other pulsars with WD companions, PSR~J1141$-$6545 and PSR~B2303+46 were not spun-up by mass transfer: they still have the large magnetic field strengths typical of young pulsars, as inferred from their spin evolution. Also, the pulsar spin axes, which are expected to have started at a random orientation with respect to the orbital plane after the SN explosion, were hence not aligned with the orbital angular momentum by an accretion process. For a compact system, such a misalignment can result in observable relativistic spin precession of the pulsar \cite{Barker&Oconnel1975}. This has been observed in PSR~J1141$-$6545 as precession led to temporal evolution of the pulse profile, providing important constraints on the system's geometry \cite{HotanEtAl2004,VenkatramanKrishnanEtAl2018}.

PSR~J1141-6545 has been observed since 2000, allowing the determination of several PK parameters including the advance of periastron, relativistic time dilation, gravitational wave damping and the Shapiro delay. These are well in agreement with GR \cite{SOM}, therefore justifying our assumption of the theory. We seek the measurement of an additional PK parameter, the temporal evolution of the observed projected-semi major axis ($x_{\rm obs}$), which to necessary precision can be written as $x_{\rm obs}\,=\,(a_{\rm p}\, \sin i / c) + A$, where $a_{\rm p}$ is the semi-major axis of the pulsar's orbit, $i$ its inclination,  $c$ is the speed of light and $A$ is the first ``aberration" parameter, which describes how the aberration of the pulsar signal affects our measurement of $x$ \cite{DamourTaylor1992,SOM}. 

Timing observations of PSR~J1141$-$6545 have been undertaken using the 64-m Parkes radio telescope and \textsc{utmost} telescopes \cite{BailesEtAl2017}. The data recording and TOA extraction followed standard pulsar data acquisition and reduction methods \cite{SOM}. The timing data were analysed using the DDGR model \cite{DDGR}, which describes the timing of the pulsar using GR. The measured and derived parameters of the system are provided in Table 1.

We measure the temporal evolution of $x$ for this system, $\dot{x}_{\mathrm{obs}} = (1.7 \pm\,0.3) \times 10^{-13}$~s\,s$^{-1}$. This value may include contributions from different physical and geometric effects depending on whether there is a corresponding change in $a_{\rm p}$, $i$ or $A$. We find only two appreciable contributions to $\dot{x}_{\mathrm{obs}}$: the largest is a change in $i$ that is due to the precession of the orbital plane caused by the spin of the WD ($\dot{x}_{\rm SO}$); with a smaller contribution arising from a change in $A$ caused by geodetic precession of the pulsar \cite{DamourTaylor1992}. The magnitude of the latter contribution was computed from the precessional constraints on the system's geometry \cite{VenkatramanKrishnanEtAl2018,SOM}: we find that it contributes $<\,21\,\%$ of $\dot{x}_{\rm obs}$ with 99\,\% confidence. The reminder is caused by $\dot{x}_{\rm SO}$, the largest contribution, which corresponds to an average increase of $i$ of 1.7 arc seconds per year. All other contributions are several orders of magnitude smaller \cite{SOM}.

Both QPM and LT effects induced by the WD spin (and only these effects) provide non-negligible contributions to $\dot{x}_{\rm SO}$:  QPM contribution ($\dot{x}_{\rm QPM}$) is inversely proportional to the square of the WD spin period ($P_{\rm WD}$); while LT contribution ($\dot{x}_{\rm LT}$) is inversely proportional to $P_{\rm WD}$. These effects also contribute to $\dot{\omega}$, but for this system the contribution is expected to be smaller than our observational uncertainties. The LT contribution to $\dot{\omega}$ might be detectable in compact double NS systems such as the double pulsar, PSR~J0737$-$3039A, in the near future \cite{KehlEtAl2016}.

Both $\dot{x}_{\rm QPM}$ and $\dot{x}_{\rm LT}$ are modulated by the spin misalignment angle ($\delta_{\rm c}$) and the precession phase ($\Phi^0_{\rm c}$; see Figure~1) of the WD at our reference epoch ($T_0$; see Table 1). Both of these angles are unknown, so we perform Markov Chain Monte Carlo (MCMC) computations to obtain a distribution for the individual contributions, and use Bayesian statistics to marginalise over the parameter space of $\delta_{\rm c}$ and $\Phi^0_{\rm c}$. From this we infer the maximum allowable $P_{\rm WD}$ consistent with the observed $\dot{x}_{\mathrm{obs}}$ \cite{SOM}.

Figure~2 shows the absolute ratio of the contributions from $\dot{x}_{\rm LT}$ and $\dot{x}_{\rm QPM}$, $R= \left| \dot{x}_{\rm LT}/\dot{x}_{\rm QPM}\right|$, as a function of $P_{\rm WD}$ (Figure~S2 shows a full correlation plot). This demonstrates that we can constrain $P_{\rm WD}\, < \, 900 \, \rm s$ to 99\% confidence. For known isolated WDs, spin periods are known to range from a few hours to a few tens of hours \cite{Kawaler2015,HermesEtAl2017}; the fastest rotating isolated WD known (SDSS~J0837+1856), which also has a mass similar to the WD in the PSR~J1141$-$6545 system ($\sim 0.9~M_{\odot}$), has spin period of $\sim\, 1.13$ hours \cite{HermesEtAl2017}. Our upper limit on $P_{\rm WD}$ is thus a confirmation of WD spin-up due to an earlier episode of mass transfer. If $P_{\rm WD} > 270$ seconds, LT is the dominant contributor to $\dot{x}_{\mathrm{SO}}$. $R$ never reaches zero (see also Figure~S2), so for all allowed values of \{$\delta_{\rm c}$,\,$\Phi^0_{\rm c}$\}, $\dot{x}_{\rm LT}$ never vanishes. Thus we detect the action of LT drag in the motion of this binary pulsar.

Taking these results as confirmation of the evolutionary history discussed above, we use binary evolution simulations to constrain $\delta_{\rm c}$ and place further constraints on $P_{\rm WD}$ \cite{SOM}. We find that the mass-transfer phase lasts for $\sim 16\,000$ years before the resulting pulsar progenitor star undergoes an ``ultra-stripped" supernova event \cite{Tauris+2013,Tauris+2015,De+2018}. If the mass-accretion rate of the $\sim 1.02$ solar mass ($M_\odot$) WD is restricted by the maximum rate of accretion before photon pressure blocks further accretion, i.e. the Eddington limit ($\sim 4\times 10^{-6}\, M_\odot\,{\rm yr}^{-1}$ for the WD), it would accrete $\sim0.06\, M_\odot$ in this time. We choose an initial orbital period and mass of the pulsar progenitor star to reproduce the most probable pre-SN binary parameters, using 70~million simulations of post-SN orbital parameters of systems resembling PSR~J1141$-$6545 \cite{SOM}.

These simulations allow us to estimate a lower limit on $P_{\rm WD}$ of $\sim$20\,s, although this depends on the interactions between the accreted material and the magnetosphere of the WD, which is not known. The angular velocity at which the WD would break up provides a firmer lower limit for $P_{\rm WD}$ of 7 s. The value of $\delta_{\rm c}$ obtained from simulations is $<50^\circ$ at 99\% confidence. This means that the WD spin is prograde, i.e. still rotating in the same direction as the orbit (before the SN the WD spin was most probably aligned with the orbit, $\delta_c \sim 0^\circ$, caused by accretion of matter from the pulsar's progenitor). We use the distribution of $\delta_c$ from simulations as a prior for our MCMC computations, obtaining tighter constraints for the distribution of $R$ and $P_{\rm WD}$ shown in Figure~2 (see also Figure~S3 for a full correlation plot). We find that $P_{\rm WD}$ is $<200 \rm \, s$ at 99\% confidence. This is because, for $\delta_{\rm c}\,<\,50^\circ$, $\dot{x}_{\rm QPM}$ is positive while $\dot{x}_{\rm LT}$ is negative. To obtain the net positive $\dot{x}_{\rm SO}$ that we observe, the WD needs to spin substantially faster so the excess from QPM ($\dot{x}_{\rm QPM} - \dot{x}_{\rm SO}$) compensates for the negative $\dot{x}_{\rm LT}$. Table 2 provides the 68\% confidence limits on $R$ and $P_{\rm WD}$ with and without binary evolution simulations. These WD spin constraints correspond to an angular momentum between  $2$--$20\times 10^{48}~ \rm g~cm^{2}~s^{-1}$.
This is one to two orders of magnitude larger than the range observed among the recycled pulsars in double NS systems, $0.03$--$0.4 \times 10^{48}~ \rm g~cm^{2}~s^{-1}$, which also likely experienced accretion onto the first-formed NS \cite{Tauris+2015}.

In summary,  measurement of the relativistic effects in the PSR~J1141$-$6545 system have enabled us to determine the masses of its WD and NS components, the orbital inclination and its variation. Only the Newtonian quadrupole spin-orbit coupling and LT precession caused by the rapidly-spinning WD predominantly contribute to this variation. LT precession is required for any orbital orientation, and is the dominant term if $P_{\rm WD}\,>\,270\, s$, after marginalising over the system's geometry. For prograde rotation of the WD, which is indicated by binary evolution simulations, $P_{\rm WD}\, < \, 200\, \rm s$ and LT precession has an opposite sign to the quadrupolar term. PSR~J1141$-$6545 is thus a unique binary pulsar that has exhibited yet another manifestation of Einstein's theory of relativity, Lense-Thirring frame-dragging.

\clearpage

\begin{center}
  \includegraphics[scale=0.8, trim=90 0 0 0,clip]{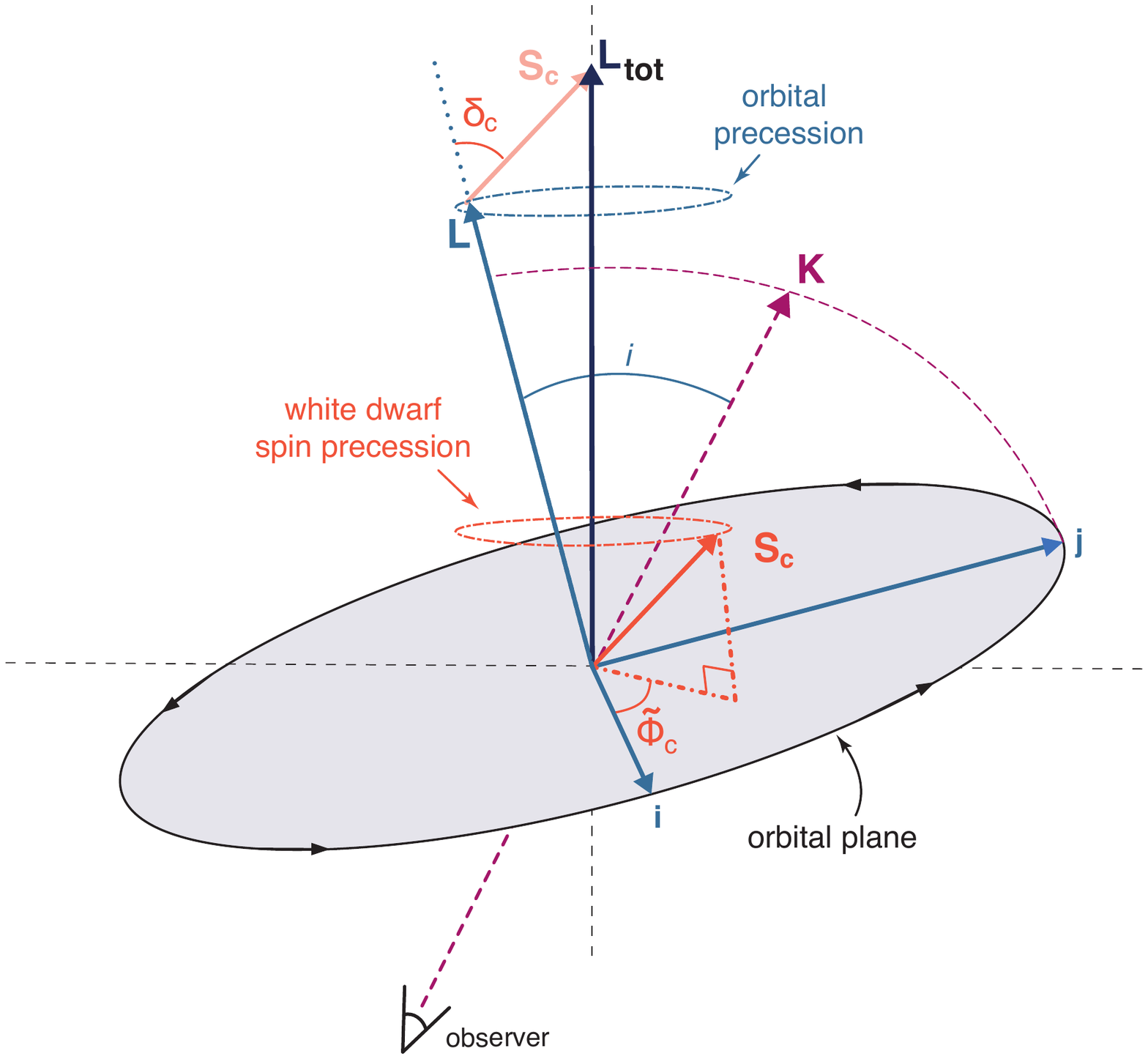} 
\end{center}
{\bf Figure~1.} {\bf Definition of the orbital geometry.} Diagram illustrating the orbital geometry of the system following the ``\textsc{DT92}" convention\cite{DamourTaylor1992}; further details are shown in Figure~S1. ${\bf L}$ is the angular momentum of the orbit, which is perpendicular to the orbital plane and inclined at an angle {\it i} to the line-of-sight vector, {\bf K}. The plane containing the vectors  ${\bf L}$ and {\bf K} intersects the orbital plane, defining the orbital plane's unit vector {\bf j} and its perpendicular counterpart,  {\bf i}. ${ \bf S_{\rm c}}$ is the spin angular momentum of the WD companion, which is misaligned from ${\bf L}$ by an angle $\delta_{\rm c}$. The vector sum of ${\bf L}$ and ${ \bf S_{\rm c}}$ forms the total angular momentum vector ${\bf L}_{\rm tot}$ which is invariant whereas  ${\bf L}$ and ${ \bf S_{\rm c}}$ precess. $\tilde{\Phi}_{\rm c}$ is the angle that the projection of ${ \bf S_{\rm c}}$ on the orbital plane subtends with respect to {\bf i}, and is related to the precession phase of the WD ($\Phi_{\rm c}$) as $\tilde{\Phi}_{\rm c}$ = $\Phi_{\rm c}$ $-270{^\circ}$. The precessions of ${\bf L}$ and ${ \bf S_{\rm c}}$ form precession cones around ${\bf L}_{\rm tot}$ as labelled. The precession of ${ \bf S_{\rm c}}$ causes $\Phi_{\rm c}$ to sweep through 360$^\circ$ whereas its rate of advance is modulated by the precession of ${\bf L}$, which induces small oscillations to the position of {\bf j} and hence in {\bf i}. Some angles and vector magnitudes in the figure are exaggerated for clarity. In practice $|\bf L| \gg |\bf S_{\rm c}|$; even if the WD is spinning at its break-up speed, the angle between $\bf L$ and $ \bf L_{\rm tot}$ is at most 0.74$^\circ$. A more detailed version can be found in Figure~S1 in \cite{SOM}.

\clearpage

\begin{center}
 \includegraphics[scale=0.63,trim=4 4 4 4,clip]{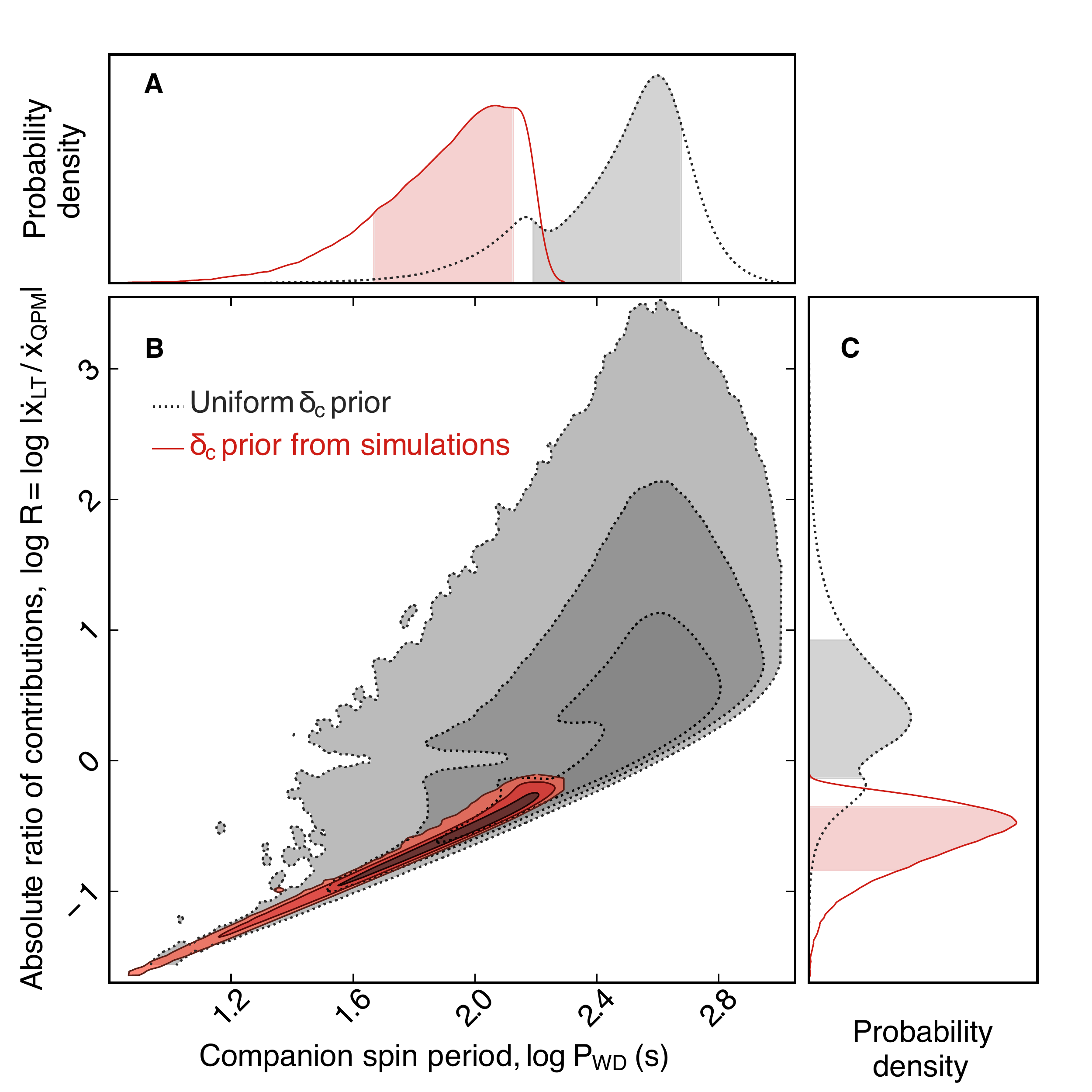} 
\end{center}
{\bf Figure~2.} {\bf Contributions to orbital precession from WD rotation.} The absolute ratio of the contributions to $\dot{x}_{\mathrm{SO}}$ from $\dot{x}_{\mathrm{LT}}$ and $\dot{x}_{\mathrm{QPM}}$, $R=\left|\dot{x}_{\rm LT}/\dot{x}_{\rm QPM}\right|$ is plotted as a function of $P_{\rm WD}$. Panel B shows the 2D probability distribution with contours defining the 68\%, 95\% and 99\% likelihood confidence intervals. The gray shaded regions and dotted contours are constraints using only the radio observations of the pulsar, while the red regions and solid contours include additional binary evolutionary constraints from simulations \cite{SOM}. Panels A and C show the marginalised posterior distributions with their 68\% confidence intervals shaded, defined by the 34\% confidence regions on either side of the 2-D maximum of the likelihood function. Numerical values are provided in Table 2.

\begin{center}
\begin{table}
\caption*{{\bf Table 1.} {\bf Model parameters for PSR~J1141$-$6545.} Shown are the post-fitting model parameter values for PSR~J1141$-$6545 with the DDGR timing model.}
\begin{tabular}{ll}
\hline\hline
\multicolumn{2}{c}{Dataset and model fit quality} \\
\hline
Modified Julian Date (MJD) range\dotfill & 51630.8 to 58214.5 (18.03 years) \\ 
Number of TOAs\dotfill & 20861 \\
Weighted root mean square timing residual ($\mu s$)\dotfill & 95.6 \\
Reduced $\chi^2$ value \dotfill & 1.0004 \\
\hline
\multicolumn{2}{c}{Fixed Quantities} \\ 
\hline
Reference epoch (MJD)\dotfill & 54000 \\ 
Glitch epoch (MJD)\dotfill & 54272.7 \\ 
\hline
\multicolumn{2}{c}{Measured quantities} \\ 
\hline
Right ascension, $\alpha$ (J2000)\dotfill &  $11^h41^m07.007^s \pm 0.003^s$ \\ 
Declination, $\delta$ (J2000)\dotfill & $-65^{\circ}45'19.14'' \pm 0.1''$ \\ 
Pulse frequency, $\nu$ (s$^{-1}$)\dotfill & $2.5387230404 \pm 1 \times 10^{-10}$ \\ 
First derivative of pulse frequency, $\dot{\nu}$~(s$^{-2}$)\dotfill & $-2.76800\times 10^{-14} \pm 1 \times 10^{-19}$ \\ 
Dispersion measure, DM (pc~cm$^{-3}$)\dotfill & $115.98 \pm 0.03$ \\ 
Orbital period, $P_{\rm b}$ (d)\dotfill & $0.19765096149 \pm 3 \times 10^{-11}$ \\ 
Epoch of periastron, $T_0$ (MJD)\dotfill & $53999.9960283 \pm 2 \times 10^{-7}$ \\ 
Projected semi-major axis of orbit, $x$ (s)\dotfill & $1.858915 \pm 3 \times 10^{-6}$ \\ 
Longitude of periastron, $\omega_0$ (degrees)\dotfill & $80.6911 \pm  6 \times 10^{-4}$\\ 
Orbital eccentricity, $e$\dotfill & $0.171876\pm  1 \times 10^{-6}$ \\ 
First derivative of $x$, $\dot{x}$ (s s$^{-1}$) \dotfill & $(1.7 \pm 0.3 ) \times 10^{-13}$ \\ 
First derivative of $e$, $\dot{e}$ (s$^{-1}$) \dotfill & $(-2 \pm 8) \times 10^{-15}$ \\ 
Companion mass, $M_{\rm c}$ ($M_\odot$)\dotfill & $1.02 \pm 0.01$ \\ 
Total Mass, $M_{\rm TOT}$ ($M_\odot$)\dotfill & $2.28967 \pm 6 \times 10^{-5} $\\ 
Glitch Phase\dotfill & $ 1.0011 \pm 0.0001$ \\ 
Glitch induced step change in $\nu$ (Hz)\dotfill & $(1.49508 \pm 0.0001)\times 10^{-6}$ \\ 
Glitch induced step change in $\dot{\nu}$ (Hz s$^{-1}$)\dotfill & $-(8.7 \pm 0.2)\times 10^{-17}$ \\ 
\hline
\multicolumn{2}{c}{Derived quantities} \\
\hline
Pulsar mass, $M_{\rm p}$ ($M_\odot$) \dotfill & $1.27 \pm 0.01$ \\
Orbital inclination, $i$ (deg) \dotfill & $71 \pm 2$ or $109 \pm 2$\\
\hline
\end{tabular}
\end{table}
\end{center}
\clearpage

\begin{center}
\begin{table}
    \centering
    \caption*{{\bf Table 2. Confidence intervals (68\%) from Figure 2.}  Shown are the 68\% confidence intervals of the companion spin period and the absolute ratio of contributions to $\dot{x}_{\mathrm{SO}}$ from $\dot{x}_{\mathrm{LT}}$ and $\dot{x}_{\mathrm{QPM}}$}
    \label{tab:model_posteriors}
    \begin{tabular}{ccc}
        \hline
        \hline
        \rule{0pt}{4ex}    
    Parameters & Uniform $\delta_{\rm c}$ prior & $\delta_{\rm c}$ prior from simulations\\
    \hline
    \hline
     \rule{0pt}{4ex}    
    Companion spin period [$P_{\rm WD}$\,(s)] & $397^{+78}_{-242}$ & $116^{+17}_{-70}$  \\
    \rule{0pt}{4ex}    
    Absolute ratio of contributions to $\dot{x}_{\mathrm{SO}}$ [R = $\left|\dot{x}_{\rm LT}/\dot{x}_{\rm QPM}\right|$]& $2.13^{+6.18}_{-1.41}$ & $0.33^{+0.11}_{-0.19}$ \\
    \hline
    \end{tabular}
\end{table}
\end{center}
\clearpage


\paragraph*{Acknowledgements}

We thank the referees and the editor for a thorough reading of the manuscript and for suggesting helpful improvements; J. Antoniadis, M. Kramer, L. Lentati, D. Reardon, R. M. Shannon, and S. Stevenson for discussions on this paper; N. Langer for the use of his binary evolution (BEC) code; L. Toomey, J. Hurley and E. Ali for help with the data release. Data reduction and analysis were performed on the gSTAR and OzSTAR national supercomputing facilities at Swinburne University of Technology. This research has made extensive use of NASA’s Astrophysics Data System (\url{https://ui.adsabs.harvard.edu/}) and includes archived data obtained through the CSIRO Data Access Portal (\url{http://data.csiro.au}) {\bf Funding:} This research was primarily supported by the Australian Research Council Centre of Excellence for All-sky Astrophysics (CAASTRO; project number CE110001020). The gSTAR and OzSTAR supercomputers are funded by Swinburne and the Australian Government's Education Investment Fund. V.V.K., N.W. and P.C.C.F. acknowledge continuing support from the Max Planck Society. M.B., C.F. and S.O. acknowledge the Australian Research Council grants OzGrav (CE170100004) and the Laureate fellowship (FL150100148). P.C.C.F. acknowledges financial support by the European Research Council for the European Research Council (ERC) starting grant BEACON, contract no. 279702. T.M.T. acknowledges an AIAS--COFUND Senior Fellowship funded by the European Union’s Horizon~2020 Research and Innovation Programme (grant agreement no~754513) and Aarhus University Research Foundation. P.A.R. acknowledges the support from the Australian Research Council (Discovery Project no. DP140102578). N.D.R.B acknowledges support from a Curtin Research Fellowship (CRF12228). The Parkes radio telescope is funded by the Commonwealth of Australia for operation as a National Facility managed by CSIRO. The Molonglo Observatory is owned and operated by the University of Sydney with support from the School of Physics and the University. {\bf Author contributions:} V.V.K. led the analysis, wrote software for data reduction, analysis and interpretation, and led writing of the manuscript. V.V.K., M.B., W.vS., and N.D.R.B. performed all the observations with the Parkes telescope. V.V.K., M.B., C.F., A.J., and S.O. performed the observations with the UTMOST telescope. V.V.K, W.vS. and S.O. performed robust polarisation calibration and profile evolution modelling. V.V.K, E.F.K. and P.A.R. performed bootstrap ToA analysis and red noise modelling. V.V.K., M.B., N.W. and P.C.C.F. interpreted the results and analysed precessional contributions to the observed value. T.M.T. performed and interpreted the binary evolution and supernova simulations. {\bf Competing interests:} The authors declare no competing interests. {\bf Data and materials availability:}Our observational data and analysis software,
including links to the software dependencies and each observational dataset, are
available at Zenodo\cite{venkatraman_krishnan_vivek_2019_3555380} 

\section*{Supplementary Materials:}

Materials and Methods\\
Supplementary Text\\
Figures S1 to S11\\
Table S1\\
References (29$-$64)

\clearpage


\setcounter{page}{1}

\def\theequation{S\arabic{equation}}

\begin{center}
Supplementary materials for \\
\vspace{\baselineskip}
\textbf{ \large Lense-Thirring frame dragging induced by a fast rotating \\white dwarf in a binary pulsar system}\\
\vspace{\baselineskip}
V. Venkatraman Krishnan, 
M. Bailes, 
W. van Straten,
N. Wex,
P. C. C. Freire, 
E. F. Keane,\\
T. M. Tauris, 
P. A. Rosado,
N.D.R. Bhat,
C. Flynn, 
A. Jameson,
S. Os{\l}owski\\
Correspondence to: vkrishnan@mpifr-bonn.mpg.de
\end{center}
\vspace{\baselineskip}

\textbf{This PDF includes}:
\begin{itemize}
  \setlength\itemsep{0.2em}
    \item[] \indent Materials and Methods
    \item[] \indent Figs. S1 to S11
    \item[] \indent Table S1
    
\clearpage    
\end{itemize}
\section*{Materials and Methods}

\beginsupplement

\paragraph*{Data recording}
The data from the Parkes telescope were taken with the central beam of the Parkes 21 cm ``multibeam" receiver  \cite{Staveley-SmithEtAl1996} using 6 different backends over the years  2000-2018. The backends used were the Analog Filterbank System (AFB), Caltech Parkes Swinburne Recorder 2 (CPSR2), three Parkes Digital Filterbanks (PDFB1, PDFB2, PDFB3) and The Collaboration for Astronomy Signal Processing and Electronics Research- CASPER Parkes Swinburne Recorder (CASPSR) (see \cite{PPTA_DR1} for details of the backends). The data from the UTMOST telescope were recorded using the The Molonglo Pulsar Swinburne Recorder (MOPSR) backend \cite{BailesEtAl2017}.  The data recording used the  \textsc{dspsr} \cite{dspsr} and \textsc{psrchive} \cite{HotanEtAl2004} software packages which in turn used the  \textsc{tempo2}\cite{EdwardsEtAl2006} pulsar timing analysis software to obtain phase predictors to fold the data at the topocentric period of the pulsar. 

Figure~S4 provides an overview of the data reduction pipeline for Parkes. The data reduction for PSR J1141$-$6545 was performed using two different polarisation calibration techniques: the Measurement Equation Template Matching (METM) technique \cite{vanStraten2013}  and the Invariant Interval \cite{Britton2000} to check for consistency between the ToAs. The millisecond pulsar PSR J0437$-$4715(Right Ascension= $04^h37^m15.8^s$, Declination = $-47\degree15'09.1''$; Epoch=J2000) was used as the polarisation reference source for METM calibrations. The data from PSR J1141$-$6545 and PSR J0437$-$4715  were first integrated up to $T_{\rm int}$ seconds (usually 180 seconds) and subjected to a median radio frequency interference (RFI) filter. Reference noise-diode observations for the flux and polarisation calibration were passed through a calibration filter which used a baseline estimation algorithm to filter out RFI prone calibration observations. The data were then flux calibrated using observations of the Hydra radio galaxy (Right Ascension=$09^h18^m05.651^s$, Declination=$-12\degree05'43.99''$, Epoch=J2000) and then polarisation calibrated. The Invariant Interval data were obtained from the METM calibrated data by taking the invariant component ($I_{inv} = \sqrt{I^2 - (Q^2 + U^2 + V^2)}$) where $I,Q,U,$ and $V$ are the Stokes parameters \cite{Lorimer&Kramer2005} of the pulse profile \cite{Lorimer&Kramer2005}, after accounting for accurate rotation measure (RM) corrections. Temporally evolving analytical standard templates were then obtained for each calibration model using an empirical profile evolution model. These standard templates were then used to obtain the Times of Arrival (ToA) of the pulses. The ToAs from the calibration and the Invariant Interval technique were cross checked and were found to agree within the uncertainties. The rest of the analysis was then performed with the METM ToAs. The ToAs were then subjected to a Bayesian red-noise and pulsar parameter estimator (\textsc{temponest}; \cite{LentatiEtAl2014}) from which estimates of pulsar parameters were obtained. The pulsar parameters also included relevant parameters for a glitch in its rotation that occured at MJD$\sim54272.7$, also reported in \cite{ManchesterEtAl2010}. Major steps in the pipeline are detailed below. The pipeline for UTMOST data reduction is provided in \cite{JankowskiEtAl2019}.

\paragraph*{Polarimetric Calibration}

Incomplete polarimetric calibration usually results in a systematic change to the total intensity profile \cite{vanStraten2004}. In the case of PSR J1141$-$6545, it also resulted in contamination of the orbital parameters, given the $\sim 4.8$-hour orbital period of the system spanning a wide range of parallactic angles in the usual full-orbit observing modes undertaken at the Parkes radio telescope. To mitigate such contaminations, polarimetric calibration was carried out using the METM technique \cite{vanStraten2013} that uses a combination of measurement equation modelling \cite[MEM]{vanStraten2004} and matrix template matching \cite[MTM]{vanStraten2006}. This technique used regular observations of a millisecond pulsar, PSR J0437$-$4715 over a wide range of parallactic angle to obtain the instrumental response of the 20-cm multibeam receiver and its variations over time. Only observations that had corresponding robust calibration solutions (with a reduced $\chi^2<1.2$) were chosen for further analysis. We used the \textsc{psrpl} data reduction pipeline which is part of the \textsc{psrchive} package \cite{HotanEtAl2004} for this analysis.

\paragraph*{Evolving pulsar profiles}

Temporal evolution of the pulse profile width and amplitude meant that we could not use a single standard template to obtain ToAs, as this could result in systematic long-term timing drifts that are co-variant with the physical parameters of interest. To combat this, we produced temporally evolving standard profiles parameterized by a set of von Mises distributions (cyclic-gaussian distributions; hereafter ``components'' ).

Firstly, the number of components required to obtain a good approximation to the observed pulse profile was estimated. To do so, the observing epoch with the widest pulse profile was chosen and fit with a set of scaled von Mises distributions parameterized by a centroid ($k$) and a concentration ($\mu$), with its probability density function taking the form $f(x | (\mu, k))  = e^{k \, \mathrm{cos} (x - \mu)}$. This differs from the original von Mises distribution by the scale factor $(1/(2\pi I_0(k) ))$ where $I_0(k)$ is a zero-th order Bessel function. Components are iteratively added to the model until the fitting residuals were sufficiently like white-noise, as determined by the value of the on-pulse residual root-mean-square (RMS) statistic being similar to its off-pulse counterpart. For PSR J1141$-$6545, a set of 3 components ($C^i; \, \forall i = \{0,1,2\}$) was found to be a good approximation to the widest pulse profile in the dataset. The values $(\mu^i, k^i) $  and their corresponding amplitudes (heights; $h^i$), were stored as the initial model ($M_{init}$). The distribution whose centroid is closest to the flux-centroid of the total intensity profile was then chosen as the ``primary" component ($C^{0}_{init}$). The relative phase-distances of the centroids of the other components from that of the primary component $C^{0}_{init}$ ($d^i = (k^i_{init} - k^{0}_{init}) \forall i = \{1,2\} $) was then held fixed for the rest of the procedure.

For each observing epoch ($E_j$), the phase centre of the pulse ($\phi^c_j$) was obtained by convolving the observation with the primary component. The primary component was then placed at ($k^{0}_j = \phi^c_j$) while other components were introduced relative to the primary component, mediated by $d^i$. The concentration and the heights of all the components were then allowed to simultaneously vary and were fit to obtain updated values $(\mu^i_{j}, h^i_{j}) $. $k^{i}_j$ need not necessarily be $k^{i}_{init}$ as inaccuracies in the initial timing model will shift the absolute phase centre $\phi^c_j$. To prevent this change being absorbed into the profile evolution model, a new model $M_{j}$ was saved with the updated concentrations and heights $(\mu^i_{j}, h^i_{j}) $ but with the original centroids $k^i_{init}$. $M_{j}$. This was then used as the initial estimate for the next observing session $E_{j+1}$. This method assumes that $\phi^c_j$ was not modified by profile evolution itself (and may not be a good assumption for other pulsars) but given the timing precision of PSR J1141$-$6545, small changes to $\phi^c_j$ from profile evolution are negligible and are absorbed into the red noise model. 

Once the corresponding models ($M_{j}$) for all epochs ($E_j$) were recorded, the temporal evolution of $(\mu^i_{j}, h^i_{j}) \forall C^i$ was fitted with a set of $5^{th}$ degree polynomial functions $X^i=\{P^i_\mu,P^i_h\}$ to obtain a smooth evolution of each component model, so that any inaccuracies in Radio Frequency Interference (RFI) rejection or calibration skewing the pulse profile shape, did not affect the analytical standard. This empirical evolution model($X^i$), was then used to create noise-free standard profiles ($A^i_k$), separated by $50$ days each. Each observing epoch ($E_j$) was then timed with the standard profile that was closest in time to the observing epoch.

\textsc{paas} program in the \textsc{psrchive} package \cite{HotanEtAl2004} was primarily used for this analysis. It was improved to accept additional keywords in the input ``initial guess'' model file such as ``fix relative phases" to fix the centroids of the components at the same relative position with respect to the primary,  ``fit primary first" to fix the centroid of the primary component to the phase centre of the pulse before introducing other components, ``set log heights" to force the height of the components to be positive and ``return original phases" that makes the updated model return the new values for concentration and height but retain the input values for the centroids. 

\paragraph*{Estimates of red noise and pulsar parameters}

Slow pulsars such as PSR J1141$-$6545 have characteristic secular drifts in their timing residuals, thought to come from emission irregularities inherent to the pulsar. Such secular drifts, for long baseline timing analysis, can be correlated with several parameters of interest and may lead to severe underestimation of the uncertainties of the pulsar parameters. To determine the uncertainties, a simultaneous model fitting for the red noise and the pulsar parameters is necessary. We use \textsc{temponest} \cite{LentatiEtAl2014}, a Bayesian pulsar timing analysis software that uses \textsc{tempo2} \cite{EdwardsEtAl2006}, the standard pulsar timing package and \textsc{multinest} \cite{FerozEtAl2009}, a Bayesian inference tool, to perform a non-linear fit for the pulsar's parameters and a red noise model combined with white-noise modifiers per backend system. These white-noise modifiers are time-independent noise sources parameterized by two values. Firstly an uncertainty factor (EFAC), which accounts for mis-calibrated radiometer noise in the system by multiplying the ToA uncertainties ($\sigma_i$) by a constant ($E_f$). Secondly, an uncertainty addition in quadrature (EQUAD), which accounts for the high frequency tail of the red noise spectrum by adding a constant ($E_q$) in quadrature to $\sigma_i$. The corrected uncertainty on each ToA can then be given as $\hat{\sigma_i}^2 = {E_f\sigma_i}^2 + E_q^2$ (this is the definition used in \textsc{temponest} which differs from its counterpart in \textsc{tempo2}). The red noise in the data is assumed to be a stationary, stochastic signal with a power-law spectrum whose spectral density $S(f)$ is given by $S(f) \propto A_{\rm red}^2\,f^{-\alpha_{\rm red}}$ where $f, A_{\rm red}$, and $\alpha_{\rm red}$ are Fourier frequency, the red noise amplitude and the red noise spectral index, respectively.

Whilst temporal variations of the dispersion measure could also produce temporal drifts in the timing residuals, the variations are expected to be at a level that is negligible for a slow pulsar like PSR J1141$-$6545. Nevertheless, we attempted to model DM variations using the DM model defined in \textsc{temponest}. Fitting this model produced posterior pulsar parameters that are consistent with the values without the addition of this parameter, and a Bayesian Information Criterion (BIC) check strongly disfavoured the addition of this parameter ($\Delta BIC >100$). 

We use the \textsc{tt(bipm)} clock correction procedure defined by the International Astronomical Union and computed  by the Bureau International des Poids et Mesures (BIPM) and use \textsc{de436} solar system model for our computations.  
The post-fitting residuals is shown in Figure~S5 and the important parameter correlations in Figure~S6.

\paragraph*{Contributions to $\dot{x}_{\rm obs}$}

In binary pulsar systems, the observed change in the projected semi-major axis of the pulsar orbit, $\dot{x}_{\rm obs}$  can arise due to a number of physical and geometric contributions, which can be decomposed as  
\begin{equation}
\dot{x}_{\mathrm{obs}} = \dot{x}_{\rm PM} + \dot{x}_{\rm \dot{D}} + \dot{x}_{\rm GW} + \dot{x}_{\dot{\rm m}} + \dot{x}_{\rm 3^{rd}} +  \dot{x}_{\dot{\epsilon}_{\rm A}} + \dot{x}_{\rm SO}
\end{equation} 

where the contributions are from the proper motion of the system ($\dot{x}_{\rm PM}$), the changing radial Doppler shift ($\dot{x}_{\rm \dot{D}}$), gravitational wave ($ \dot{x}_{\rm GW}$) emission,  mass-loss in the system ($\dot{x}_{\dot{\rm m}}$), the presence of a hypothetical third body in the system ($\dot{x}_{\rm 3^{rd}}$), a secular change in the aberration of the pulsar beam due to geodetic precession ($\dot{x}_{\dot{\epsilon}_{\rm A}}$), and spin-orbit ($ \dot{x}_{\rm SO}$) coupling \cite{Lorimer&Kramer2005}. We examine each of these parameters in the following subsections.

\subparagraph*{Proper motion and changing Doppler shift}

The maximal contribution from $\dot{x}_{\rm PM}$ is given by 
\begin{equation}
\label{eqn:xdot_pm}
\dot{x}_{\rm PM} \leq1.54 \times 10^{-16}  x \cot i \left( \frac{\mu_{\rm T}}{\rm mas~yr^{-1}} \right)
\end{equation}
where $\mu_{\rm T}$ is the total proper motion on the sky \cite{DD2,DamourTaylor1992}. 

\noindent The contribution from the changing Doppler shift is given by 
\begin{equation}
\label{eqn:xdot_doppler}
\dot{x}_{\rm \dot{D}} \sim  x \left[\left(\frac{V_{\rm T}^2}{dc}\right)+ \frac{\vec{K_0}\cdot (\vec{a}_{\rm PSR} - \vec{a}_{\rm SSB})}{c}\right]
\end{equation}
where $d$ is the distance to the pulsar and $V_{\rm T} = \mu_{\rm T}d$ is its corresponding transverse velocity, $\vec{K_0}$ is the unit vector from the Solar system barycentre to the pulsar and $(\vec{a}_{\rm PSR} - \vec{a}_{\rm SSB})$ is the differential Galactic acceleration of the pulsar with respect to the Solar system barycentre \cite{DD2,DamourTaylor1992}.

We estimated the contributions of $\dot{x}_{\rm PM}$ and $\dot{x}_{\dot{\rm D}}$ using a variety of measurements of the proper motion and the distance to the pulsar. Estimates of the proper motion were obtained from scintillation velocity measurements \cite{Ord2002_SV,ReardonEtAl2018}. Firstly, the dispersion measure (DM) of the pulsar was used along with two different Galactic electron density models (namely the NE2001 \cite{Cordes&Lazio2002} and YMW16 \cite{YMW16} ) to obtain distance estimates of 2.4 kpc and 1.6 kpc respectively. The neutral hydrogen absorption estimate of the lower limit to the distance is 3.7 kpc \cite{Ord2002_NH}. A distance estimate from scintillation velocity measurements is $\sim 10 ^{+4}_{-3}$ kpc \cite{ReardonEtAl2018}. Although these estimates are inconsistent, the maximum possible contributions of $\dot{x}_{\rm PM}$ and $\dot{x}_{\dot{\rm D}}$, by taking the most conservative distance estimate, was $\sim 2\%$ of $\dot{x}_{\rm obs}$ at most for each.

\subparagraph*{Gravitational wave emission}

We estimate the $\dot{x}_{\rm GW}$ contribution to $\dot{x}_{\rm obs}$ using the measured rate of change of the orbital period ($\dot{P}_{ \rm b}^{\rm obs}$) which, for this exercise can be assumed to be the contribution to $\dot{P}_{ \rm b}$ from gravitational wave emission ($\dot{P}_{ \rm b}^{\rm GW}$), since

\begin{equation}
\label{eqn:pbdot_xdot}
   \frac{\dot{x}^{\rm GW}}{x} = \frac{2}{3}\,\frac{\dot{P}_{\rm b}^{\rm GW}}{P_{\rm b}}
   \simeq \frac{2}{3}\,\frac{\dot{P}_{\rm b}^{\rm obs}}{P_{\rm b}} \,.
\end{equation}

\noindent We find it to be of the order of $10^{-18} \rm \,s~s^{-1}$ . This is 5 orders of magnitude smaller than  $\dot{x}_{\rm obs}$. This term is hence negligible and is ignored in this analysis.

\subparagraph*{Mass loss in the system}

A mass loss from the system due to radiation emission from the neutron star and/or a wind from the companion would result in an additional contribution to the orbital period derivative,  $\dot{P}_{\rm b}^{\rm obs}$. A limit for the contribution to  $\dot{x}_{\rm obs}$ from mass loss in the system ($\dot{x}_{\rm \dot{m}}$) can then be obtained by using the residual observational uncertainty ($\Delta \dot{P_{\rm b}}$) on $\dot{P}_{\rm b}^{\rm obs}$ after subtracting the contribution from $\dot{P}_{\rm b}^{\rm GW}$  (see \cite{DamourTaylor1991}, their eqs.~(4.1) and (4.2) ) 
\begin{equation}
    \label{eqn:massloss_xdot}
    \frac{\dot{x}_{\rm \dot{m}}}{x} = \frac{\dot{a}_{\rm \dot{m}}}{a} \sim \frac{\Delta\dot{P}_{\rm b}}{P_{\rm b}},
\end{equation}
where $a$ denotes the semi-major axis of the relative orbit, and $a_{\dot{\rm m}}$ its change due to mass loss. We find this contribution to be of of the order of $10^{-19} \rm \,s~s^{-1}$ which is 6 orders of magnitude lower than our detection. This term is hence negligible and is ignored in this analysis.

\noindent

\subparagraph*{Presence of a hypothetical third body}

We rule out the presence of any hypothetical third body in our system as it should have resulted in contributions to the higher order frequency derivatives of the pulsar spin. Including higher order spin derivatives in our model fitting returned posterior distributions consistent with 0 and were also strongly disfavoured ($\Delta BIC >20$).

\subparagraph*{Rate of change of aberration}

The orbital motion of the rotating pulsar causes the pulsar beam to be ``aberrated" into a distant observer's line of sight. The effects of aberration on pulsar timing are not separately measurable as they are completely absorbed as a redefinition of the Keplerian parameters \cite{DD2}. Such redefinitions, among others, also cause the observed projected length of the semi-major axis ($x_{\rm obs}$) to differ from the intrinsic value ($x_{\rm intrinsic}$) by a factor of (1 + $\epsilon_{\rm A}$), where $\epsilon_{\rm A}$ is the first aberration parameter ($A$) divided by $x$ \cite{DamourTaylor1992}. This aberration term depends on the pulsar spin period, the Keplerian parameters, and the system's polar angles. In a spin-aligned system, this would mean that aberration can never be distinctly measured using pulsar timing alone. However, for misaligned systems such as PSR J1141$-$6545, geodetic precession gives rise to changes to the geometry of the system, and hence changes $\epsilon_{\rm A}$ on timescale of the geodetic precession. This causes an apparent secular evolution of Keplerian parameters including the orbital eccentricity and projected semi-major axis, thus contributing to $\dot{x}_{\rm obs}$.

The contribution to $\dot{x}_{\rm obs}$ from a change in the aberration due to the geodetic spin-precession of the pulsar ($\dot{x}_{\dot{\epsilon}_{\rm A}}$) is given by 
\begin{equation}
\dot{x}_{\dot{\epsilon}_{\rm A}} 
= x\left(\frac{d\epsilon_{\rm A}}{dt}\right) 
= -x\,\frac{P}{P_{\rm b}} \, \frac{\Omega_{\rm geod}}{(1- e^2)^{1/2}} \, 
  \frac{\cot \lambda_{\rm p} \sin 2\eta_{\rm p} + \cot \textit{i} \cos \eta_{\rm p}}{\sin \lambda_{\rm p}}
\end{equation}
\cite{DD2,DamourEspositoFarese1992}, where $P_{\rm b}$ is the orbital period, $\Omega_{\rm geod}$ is the geodetic precession rate, $\lambda_{\rm p}$ is the angle between the pulsar spin and the line of sight, $\eta_{\rm p}$ is the longitude of precession (see Figure~1 and S1 for definitions of these and other angles). The contribution to $\dot{x}_{\rm obs}$ from the rate of change of aberration hence depends on the system's geometry. The inclination angle was measured using the DDGR timing model as $i=71 \degree \pm 2 \degree$. An equally likely solution for $i$ arising from symmetry, $(180\degree - i) = 109 \degree \pm 2\degree$, has been ruled out using measurements of the pulsar's scintillation velocity \cite{ReardonEtAl2018}. While $\eta_{\rm p}$ mediates the sign of $\dot{x}_{\dot{\epsilon}_{\rm A}}$, its absolute magnitude increases rapidly (see Figure~S7) as $\lambda_{\rm p}$ approaches $0\degree$ (or 180$\degree$), meaning the pulsar's spin (or anti-spin) is very close to our line-of-sight. Hence it is necessary to constrain the ranges of $\lambda_{\rm p}$ and $\eta_{\rm p}$.

An analysis of the evolving total intensity pulse profile  (due to relativistic spin precession) was performed to understand the orientation and the geometry of the pulsar \cite{VenkatramanKrishnanEtAl2018}. Under the usual assumption of the circular beam model of pulsar emission, we set limits on $\lambda_{\rm p}$ of $66 \degree < \lambda_{\rm p} < 114 \degree$, marginalizing over all other geometric evolutions, with 99\% confidence \cite{VenkatramanKrishnanEtAl2018}. This shows that the contributions of ${\dot{x}}_{\dot{\epsilon}_{\rm A}}$ from aberration is at most 20\% of $\dot{x}_{\mathrm{obs}}$. There are a number of other arguments to support this conclusion. Firstly, it is unlikely that the pulsar spin (or anti-spin) that is randomly oriented on the sky is close to our line of sight. Secondly, such a close alignment of the spin axis to our line of sight would make the pulsar a nearly-aligned rotator, while still possessing a narrow duty-cycle ($\sim 10\%$) across 20 years of precession \cite{ManchesterEtAl2010,VenkatramanKrishnanEtAl2018}, which is also unlikely, as the near-alignment would require the pulsar's beam to be pointing towards the earth for most of its rotational phase. Thirdly, if we assume such a nearly-aligned orientation, and that $\dot{x}_{\rm obs}$ is entirely due to $\dot{\epsilon}_{\rm A}$, there is another observable that should be detectable: the rate of change of observed eccentricity ($\dot{e}$).  To first order,the predominant contributions to ${\dot{e}}$ come from gravitational wave emission ($\dot{e}_{\rm GW}$) and the rate of change of aberration ($\dot{e}_{\dot{\epsilon}_{\rm A}}$). Neglecting the $\sim 10^{-18}$ contribution from $\dot{e}_{\rm GW}$, the contributions from $\dot{\epsilon}_{\rm A}$ to $\dot{x}_{\rm obs}$ and $\dot{e}_{\rm obs}$ are expected to be related as ${\dot{x}}_{\dot{\epsilon}_{\rm A}}/{x} = {\dot{e}}_{\dot{\epsilon}_{\rm A}}/{e}$. Consequently, if $\dot{x}_{\rm obs}$ is entirely due to $\dot{\epsilon}_{\rm A}$, then we expect $\dot{e} \simeq (e/x)\dot{x}_{\rm obs} \simeq 10^{-14} \, \rm s^{-1}$. To test this, we re-analyzed the data with an additional model parameter for $\dot{e}$. Our measured $\dot{e}_{\rm obs}$ = $ (-2  \pm 8) \times 10^{-15} \rm \, s^{-1}$ is consistent with a non-detection, with its mean value that is below the expected detection level of $+1 \times 10^{-14}\, \rm s^{-1}$, for an $\dot{x}_{\rm obs}$ solely due to changing aberration. The large variance however, does not provide a stringent constraint on the contribution to $\dot{x}_{\rm obs}$ from $\dot{\epsilon}_{\rm A}$, although a constraint of $ {\dot{x}}_{\dot{\epsilon}_{\rm A}} < 0.4~\dot{x}_{\rm obs}$ is obtained with 68\% confidence. Additionally, a BIC check also strongly disfavors ($\Delta BIC \sim 10$) the addition of the extra model parameter. All the above arguments favour results derived from the pulsar's geodetic precession. A precession of the orbital plane due to spin-orbit interactions ($\dot{x}_{\rm SO}$) should then provide the dominant contribution ($> 79\%$) to $\dot{x}_{\rm obs}$ (see Figure~S8).

\subparagraph*{Spin-orbit interaction from the white-dwarf}

The contribution of classical spin-orbit coupling ($\dot{x}_{\rm QPM}$) due to the rotationally induced quadrupole moment of the white dwarf to $\dot{x}_{\rm obs}$ is given by: 
\begin{equation}\label{eq:QPM}
  \dot{x}_{\rm QPM} = x\left(\frac{2\pi}{P_{\rm b}}\right) Q \cot\textit{i} 
    \sin2\delta_{\rm c} \sin\Phi^0_{\rm c}
\end{equation}
where
\begin{equation}
  Q = \frac{k_2 R^2_{\rm c} \hat{\Omega}_{\rm c}^2}{2a^2(1- e^2)^2} \quad {\rm with} \quad 
 \hat{\Omega}_{\rm c} \equiv \frac{\Omega_{\rm c}}{(Gm_{\rm c}/R_{\rm c}^3)^{1/2}}
\end{equation} 
and $\Omega_{\rm c} = 2\pi/P_{\rm WD}$  \cite{SmarrBlandford1976,Lai95,Wex98}. For the radius of a $1.02\,M_\odot$ white dwarf one finds $R_c \simeq 5400\,{\rm km}$ (assuming WD to be an ideal degenerate Fermi gas \cite{Chandrasekhar1931}) . The equation-of-state and composition independent I-Love-Q relations \cite{Boshkayev2017} for the WD were used to obtain $k_2 = 0.081$. Equation \ref{eq:QPM} includes only the contributions of spin-induced quadrupole moment. Contributions from tides and magnetic fields negligible and have been ignored.

The contribution to $\dot{x}_{\rm obs}$ from Lense-Thirring precession ($\dot{x}_{\rm LT}$) due to the WD spin is given by 
\begin{equation}
 \label{eq:LT}
 \dot{x}_{\rm LT} \simeq - x\frac{GS_{\rm c}}{c^2 a^3 (1-e^2)^{3/2}}\left(2 + \frac{3m_{\rm p}}{2m_{\rm c}}\right) \cot \textit{i} \sin{\delta_{\rm c}} \sin{\Phi^0_{\rm c}} 
 \end{equation} 
 where $S_{\rm c} = I_c\Omega_{\rm c}$ is the WD angular momentum, with $I_c$  being its moment of inertia \cite{DamourTaylor1992}.

\subparagraph*{MCMC simulations of spin-orbit coupling}

We perform MCMC simulations with the \textsc{emcee} software package \cite{EMCEE} in the \textsc{Python} programming language, and solve for equations \ref{eq:QPM} and \ref{eq:LT} simultaneously, to find the minimum WD spin period that could give rise to the spin-orbit coupling contributions to $\dot{x}_{\rm obs}$, using uniform priors for the possible range of the angles $\{\delta_{\rm c}, \Phi^0_{\rm c}\}$. We use the Gelman-Rubin convergence criterion \cite{GelmanRubin1992} to check the convergence of our MCMC chains implemented in the \textsc{ChainConsumer} package \cite{Hinton2016_ChainConsumer}. We marginalise over the dimensions $\{\delta_{\rm c}, \Phi^0_{\rm c}\}$ to obtain the minimum period of the WD, regardless of its orientation, along with the individual contributions of classical quadrupole and Lense-Thirring interactions.

Using the known values for $M_{\rm c},R_{\rm c},I_{\rm c}$ and $k_2$ in \ref{eq:QPM} and \ref{eq:LT} we obtain the following simplified equations for the relative contributions from $\dot{x}_{\rm QPM}$ and $\dot{x}_{\rm LT}$:
\begin{eqnarray}
    \label{eq:simp_QPM}
    \dot{x}_{\rm QPM} = 0.88 \times 10^{-13} (P_{\rm WD}/5\,{\rm min})^{-2}\sin2\delta_{\rm c}\sin\Phi_{\rm c}^0,\\
 \label{eq:simp_LT}
    \dot{x}_{\rm LT} = -2.58 \times 10^{-13} (P_{\rm WD}/5\,{\rm min})^{-1}\sin\delta_{\rm c}\sin\Phi_{\rm c}^0.
\end{eqnarray}
It can be seen from these equations that $\dot{x}_{\rm QPM}$ and $\dot{x}_{\rm LT}$ contribute on the same order of magnitude to $\dot{x}_{\rm SO}$ if $P_{\rm WD}$ is a few minutes.

\paragraph*{Contributions to $\dot{\omega}$ from spin-orbit interaction}

The spin-orbit interactions also result in contributions to the advance of periastron, whose magnitudes depend on angles $\{\delta_{\rm c}, \Phi^0_{\rm c}\}$ \cite{Wex98}. We find that the present precision of $\dot{\omega}$ is insufficient to place any limit on the additional contribution from spin-orbit coupling and hence cannot be used to constrain the $\delta_{\rm c}-\Phi^0_{\rm c}$ parameter space. The contribution from the tidally induced quadrupole moment is at most $\sim1\%$ of the spin induced contribution \cite{Shakura1985}.
\clearpage

\paragraph*{Consistency with binary evolution}
For our calculation of the evolution of a helium star-white dwarf (WD) binary, the progenitor system of PSR J1141$-$6545, we applied the \textsc{bec} stellar evolution code, i.e. the ``Langer code'' \cite{ywl10,tlk12,ltk+14,Tauris+2013,Tauris+2015}. We assumed an initial helium star donor mass of $M_{\rm He}=2.9\,M_{\odot}$,
a WD mass of $M_{\rm WD}=0.96\,M_{\odot}$ and an orbital period of $0.40\,{\rm days}$. The WD mass and the orbital period were found by iteration until the post-Roche-lobe overflow (RLO) solution matched the present mass of the WD ($1.02\,M_{\odot}$) and a pre-SN orbital period (0.14~days) in accordance with our SN analysis reproducing the PSR J1141$-$6545 system (see below).

Figure~S9 shows the evolution of the helium-star donor in a Kippenhahn diagram, displaying the evolution of the internal structure of the star. We were unable to evolve the star until core collapse (mainly due to a rigorous helium shell flash \cite{Tauris+2013}), but the core/envelope boundary and the total mass are frozen in that remaining short epoch \cite{Tauris+2015}. At the end of the RLO, the helium star has been stripped down to a pre-SN mass of only $1.58\,M_\odot$ and a remaining helium envelope of about $0.06\,M_\odot$. These are typical characteristics for ultra-stripped SNe \cite{Tauris+2013,Tauris+2015,De+2018}. Given the relatively small mass of the resulting NS observed in the PSR J1141$-$6545 system ($1.27\,M_\odot$), we expect a low kick velocity of $\sim 50\,{\rm km\,s}^{-1}$ \cite{Tauris+2017}, producing a small misalignment angle of the WD spin axis with respect to the orbital angular momentum vector in the present system (see further discussions below on the effect of the SN).

Figure~S10 shows the calculated mass-transfer rate as a function of time (since the helium star was on its zero-age main sequence) which is highly super-Eddington. The Eddington accretion rate for a $\sim 1.0\,M_\odot$ WD accretor is $\dot{M}_{\rm Edd}\simeq 4\times 10^{-6}\,M_\odot\,{\rm yr}^{-1}$. 
The accumulated phase of Case~BB (helium shell burning) mass transfer lasts for about $16\,000\,{\rm yr}$.
Assuming the accretion onto the WD to be limited to the Eddington rate, it accretes about $\Delta M_{\rm WD}=0.06\,M_\odot$, 
which is more than sufficient to spin it up to a high spin rate, as we demonstrate below.

\paragraph*{WD accretion and spin up}
The mass transferred from the helium-star donor carries angular momentum which eventually spins up the WD.
Besides material stress, the accretion torque, $N$ acting on the WD has a contribution from both magnetic stress and viscous stress 
and its effect can be expressed as:
$N = \dot{J}_\star \equiv (d/dt) (I\Omega _\star)$,
where $J_\star$ is the WD spin angular momentum, $\Omega _\star$ is its angular velocity and 
$I=k^2\,M_{\rm WD}R_{\rm WD}^2\approx 0.9 \times 10^{50}\,{\rm g~cm}^2$ is its moment of inertia (computed from eqn. (4) of \cite{Boshkayev2017}), where $k$ is the gyration radius.

The exchange of angular momentum ($\vec{J}_\ast=\vec{r}\times\vec{p}$, where $\vec{r}$ is the position vector of a particle, and $\vec{p}$ is its momentum vector) at the magnetospheric boundary eventually leads to a
gain of WD spin angular momentum which can approximately be expressed as:
\begin{equation}\label{eq:spin-ang}
  \Delta J_\star = \int n(\omega,t)\,\dot{M}_{\rm WD}(t)\,\sqrt{GM_{\rm WD}(t)r_{\rm mag}(t)}\;\xi (t)\;dt \simeq \sqrt{GM_{\rm WD}r_{\rm mag}}\;\Delta M_{\rm WD}
\end{equation}
where $M_{\rm WD}$ denotes the WD mass, $\dot{M}_{\rm WD}$ is its mass-accretion rate, 
$n(\omega)$ is a dimensionless torque and $\xi \simeq 1$ is a numerical factor which depends on the flow pattern \cite{gl79b,gl92,tlk12}.
The total amount of mass accreted by the WD is given by $\Delta M_{\rm WD}$, and
$r_{\rm mag}$ is the location of the magnetosphere (the Alfv\'en radius, or the length of the lever arm) defined as the location where the magnetic energy density will begin to control the flow of the material (i.e. where the incoming material couples to the magnetic field lines
and co-rotate with the WD magnetosphere), given by \cite{bv91}:
\begin{equation}\label{eq:alfven}
   r_{\rm mag} \simeq \left( \frac{B^2\,R_{\rm WD}^6}{\dot{M}_{\rm WD}\sqrt{2GM_{\rm WD}}} \right) ^{2/7}. 
\end{equation}
A typical value for the magnetospheric radius of an Eddington-accreting WD (assuming $\dot{M}_{\rm Edd}=4\times 10^{-6}\,M_\odot\,{\rm yr}^{-1}$) is thus $r_{\rm mag}\simeq 1-16\,R_{\rm WD}$ for a surface magnetic flux density of $B=10^6-10^8\,{\rm G}$.
For a weaker B-field, the magnetosphere is pushed down to the surface of the WD and the lever arm of the accretion torque is simply $R_{\rm WD}$.

We can now estimate the minimum spin period, $P^{\rm min}_{\rm WD}$ of the WD in the PSR J1141$-$6545 system for the most favorable torque transmission (see Eq.~\ref{eq:spin-ang}):
\begin{equation}\label{eq:Pmin}
   P^{\rm min}_{\rm WD} = \frac{2\pi\,I}{\sqrt{GM_{\rm WD}r_{\rm mag}}\,\Delta M_{\rm WD}} \simeq 20\,{\rm sec}\;\left( \frac{R_{\rm WD}}{r_{\rm mag}} \right)^{1/2} 
\end{equation}
where we have inserted our estimate of accreted material of $\Delta M_{\rm WD}\simeq 0.06\,M_\odot$.

So far, we have disregarded the possibility of propeller effects at work in case $r_{\rm mag}>r_{\rm co}$, where the co-rotation radius is given by: 
\begin{equation}\label{eq:corot}
   r_{\rm co} \equiv \left( \frac{GM_{\rm WD}P_{\rm WD}^2}{4\pi^2} \right)^{1/3} \simeq 6\;R_{\rm WD}\;\left(\frac{P_{\rm WD}}{\rm 100\,s}\right)^{2/3}
\end{equation} for the WD companion to PSR J1141$-$6545. Hence, we conclude that propeller effects can be disregarded, unless the B-field of the WD is exceptionally strong while its spin period, $P_{\rm WD}$ is very small.

With our derived 99\% upper limits on $P^{\rm min}_{\rm WD}$ being $\sim 900$~seconds (Figure~2 and S2), we
conclude that our interpretation of the PSR J1141$-$6545 system is consistent with expectations from binary stellar evolution and standard accretion physics. Hence, we use the posterior distribution of the spin-misalignment angle $\delta_{\rm c}$ from the subsequent supernova simulations to further constrain the WD spin to $< 200$ seconds.

\paragraph*{Simulations of the supernova producing PSR J1141$-$6545}
We have examined the the kinematic effects of 70~million simulated supernova (SN) explosions to reproduce the measured orbital parameters of the PSR J1141$-$6545 system, following the method applied in \cite{Tauris+2017}. 
In our Monte Carlo simulations, we take advantage of our derived masses for the NS ($1.27\,M_\odot$) and the WD ($1.02\,M_\odot$) from observations, as well
as the measured values of orbital period (here denoted $P_{\rm b}=0.198\,{\rm days}$) and eccentricity ($e=0.172$). Given that PSR J1141$-$6545 is a young radio pulsar, we neglect any orbital evolution of the system since its formation. 
We simulate SNe over a 5-dimensional phase~space whose parameters are: the pre-SN orbital period, $P_{\rm b,0}$; the final mass of the (ultra-stripped) exploding helium star, $M_{\rm He}$; the magnitude of the kick velocity imparted on the newborn NS, $w$; and the two angles defining the direction of the kick velocity, $\theta$ and $\phi$. 
We assumed a flat non-informative prior distributions for all these parameters except for $\theta$ where a random (isotropic) kick direction leads to a prior distribution of $\sin(\theta)$. The extents of the prior probability distributions are given in Table S1.

Using Monte Carlo methods, we repeatedly select a set of values of $P_{\rm b,0}$, $M_{\rm He}$, $w$, $\theta$ and $\phi$, and solve in each trial for the post-SN orbital parameters as outlined in \cite{Tauris+2017}.
From the outcome of the initial simulations, we can compare with the values of the PSR J1141$-$6545 system and iterate by adjusting the pre-SN parameter space until the outcome matches with the observed post-SN values within a chosen error margin of 3\%. In Figure~S11, we show the results of our simulations. The chosen solutions are centered on the observed values of $(P_{\rm b},\,e)$ with the accepted error margin of $\pm$3\% in both
$P_{\rm b}$ and $e$ as shown in Figure~S11A. Figure~S11B shows the post-SN distribution of 3D systemic velocities. 
Figure~S11C and D show the pre-SN orbital period, $P_{\rm b,0}$ and the mass of the exploding star, $M_{\rm He}$.
Panel~E shows the applied kick magnitudes for all successful solutions; and Figure~S11F shows the distribution of resulting post-SN misalignment angles of the WD.

In these simulations, we restricted the mass of the exploding star to $\le 1.7\,M_\odot$ based on the mass-transfer calculations (see above), which resulted in an ultra-stripped star prior to the explosion. Given that the NS mass of PSR J1141$-$6545 is rather low ($1.27\,M_\odot$), we also restricted the kick magnitude to $\le 400\,{\rm km\,s}^{-1}$ --- an upper limit derived for the similar second SN explosion in ultra-stripped SNe producing double NS systems \cite{Tauris+2017}. The latter constraint is also supported by the possible correlation between NS mass and kick magnitude \cite{Tauris+2017,Janka2017}.

The misalignment angle of the WD is found to be less than $50^\circ$ at the 99\%  confidence level. Misalignment angles in the interval between $40^\circ - 90^\circ$ appear in 10\% of the cases only if we allow for extremely large kicks ($400-1000\,{\rm km\,s}^{-1}$), and masses of the exploding star up to $2.5-3.0\,M_\odot$ (which are disfavoured by the evolution and expansion of the progenitor helium star in binary stellar evolution models).

From additional simulations, we also find that the distribution of resulting misalignment angles is insensitive to changes in the allowed error margin in post-SN orbital period and eccentricity (e.g. applying $\pm 10$\% yields a very similar result). This is because post-SN systems with properties somewhat similar to those of PSR~J1141$-$6545, would all have fairly small SN ejecta and a small NS mass, resulting in a relatively small kick, and thus small misalignment angles.

\clearpage

\begin{center}
 \includegraphics[scale=0.3,trim=30 400 40 500,clip]{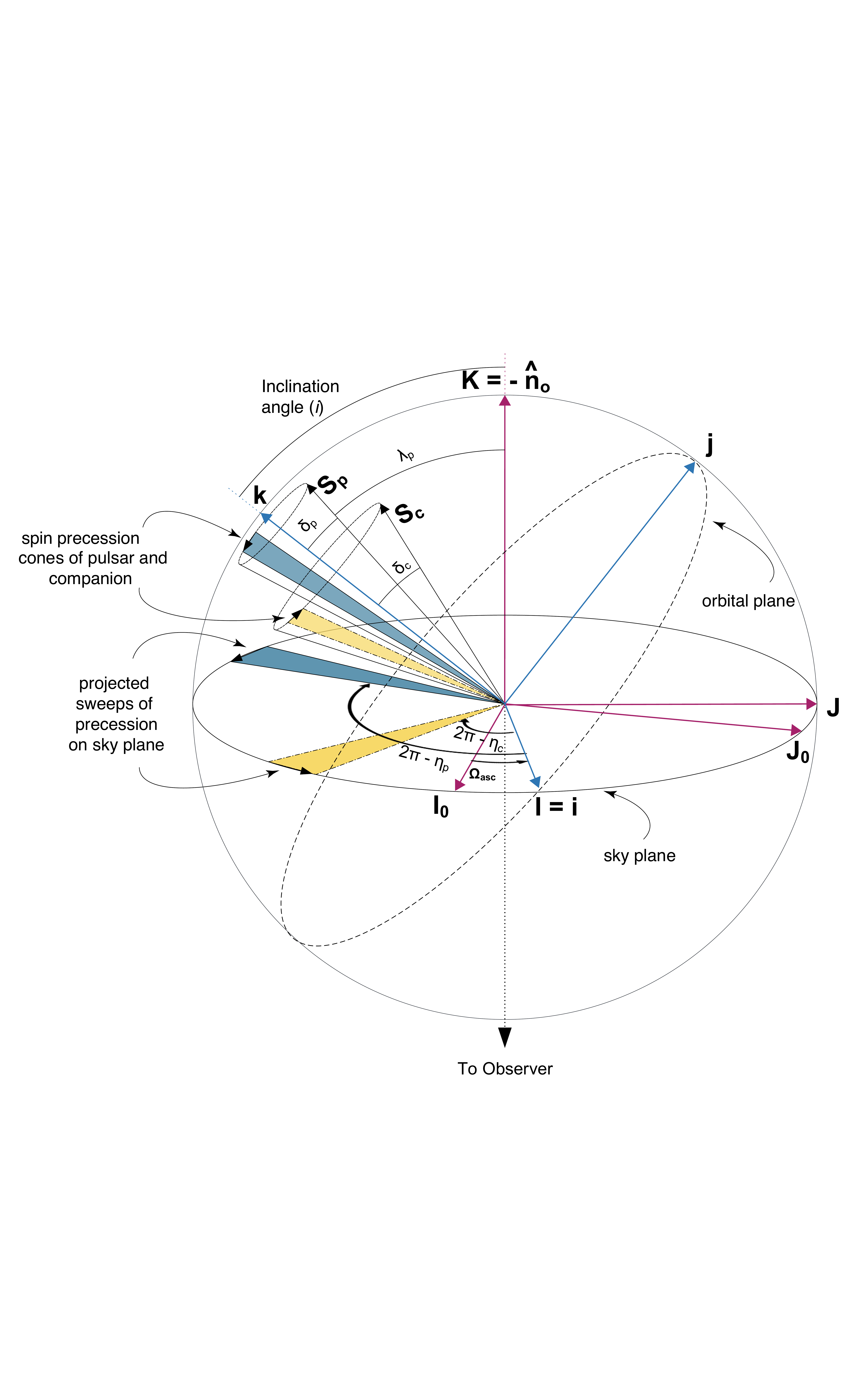} 
\end{center}
{\bf Figure~S1.} {\bf Definition of spin and orbital geometry.} Illustration of various angles in the spin and orbital geometry of the system, using the "DT92" definitions \cite{DamourTaylor1992}. The unit vectors ${\bf I}_0$  and ${\bf J}_0$ denote the plane of the sky, with {\bf K} denoting the line of sight from the observer to the pulsar, and perpendicular to the sky plane. The unit vectors {\bf I$\equiv$i} and  {\bf j} form the orbital plane, inclined at an angle {\it i} to the sky plane and rotated in azimuth by $\Omega_{\rm asc}$. The normal to the orbital plane ({\bf k}) is then by definition the direction of the orbital angular momentum vector ${\bf L}$ which is approximated here to be also the direction of the total angular momentum vector ${\bf L}_{\rm tot}$, given that the magnitude of the angular momenta of the component stars are much smaller than ${|\bf L|}$ (See Figure~1 for the case where this approximation is not made). The spin angular momentum vectors of the pulsar and the companion are given by ${ \bf S_{\rm p}}$ and ${ \bf S_{\rm c}}$ respectively, which are misaligned from ${\bf L}_{\rm tot}$ by angles  \textbf{$\delta_{\rm p}$} and \textbf{$\delta_{\rm c}$}. \textbf{$\lambda_p$} is defined as the angle between \textbf{${ \bf S_{\rm p}}$} and \textbf{K}. Spin precession causes the spin vectors (${ \bf S_{\rm p}}$, ${ \bf S_{\rm c}}$) to precess around ${\bf L}_{\rm tot}$ at a rate denoted by $\Omega_{\rm p}^{\rm geod}$ and $\Omega_{\rm c}^{\rm geod}$ respectively. This causes the instantaneous precession phases of the pulsar and the companion  to evolve in time, thereby sweeping out ``precession cones". These precessional sweeps projected onto the sky plane sweep the corresponding longitudes of precession ($\eta_{\rm p}$, $\eta_{\rm c}$), as shown by the blue and yellow shaded sweeps on the precession cones and the sky plane. These angles are measured from \textbf{I} and the complementary angles \textbf{$(2\pi-\eta_{\rm p})$} and \textbf{$(2\pi-\eta_{\rm c})$} are shown for clarity.

\clearpage

\begin{center}
  \includegraphics[scale=0.6,trim=4 10 10 10,clip]{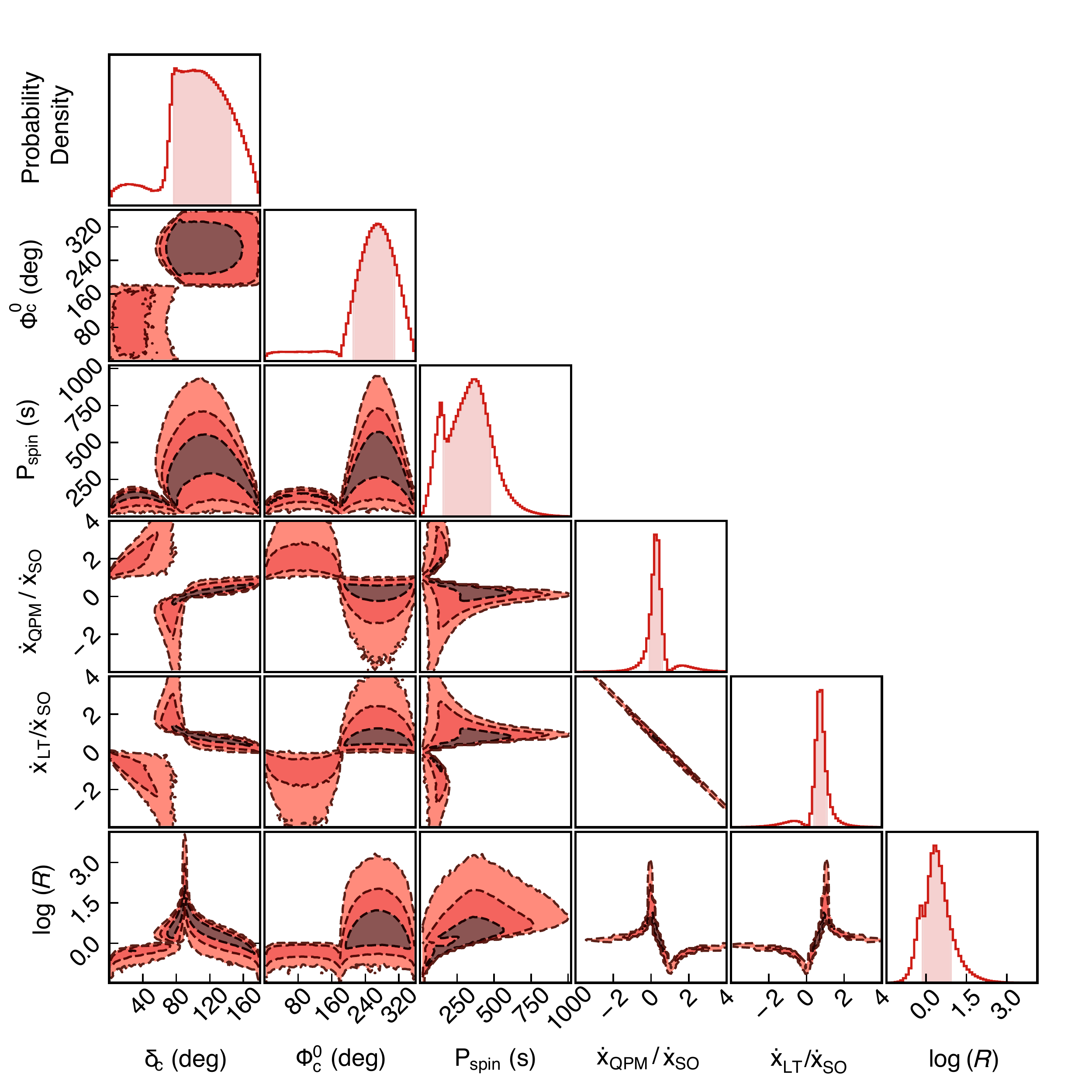} 
\end{center}
{\bf Figure~S2.} {\bf Contributions to orbital precession from WD rotation.} This correlation plot shows the fractional contribution to $\dot{x}_{\mathrm{SO}}$ from $\dot{x}_{\mathrm{QPM}}$ and $\dot{x}_{\mathrm{LT}}$ and their absolute ratio ($R$) as a function of the apriori unknown angles \{$\delta_{\rm c}$, $\Phi^0_{\rm c}$\} and $P_{\rm WD}$. The off-diagonal panels are 2D probability densities with dashed contours enclosing 68\%, 95\% and 99\% iso-likelihood confidence intervals, shaded in progressively darker colors. The diagonal probability densities with solid lines are marginalised posterior probability distributions of the corresponding dimensions, with shaded regions indicating the 68\% iso-likelihood contours.

\clearpage

\begin{center}
  \includegraphics[scale=0.6,trim=4 10 10 10,clip]{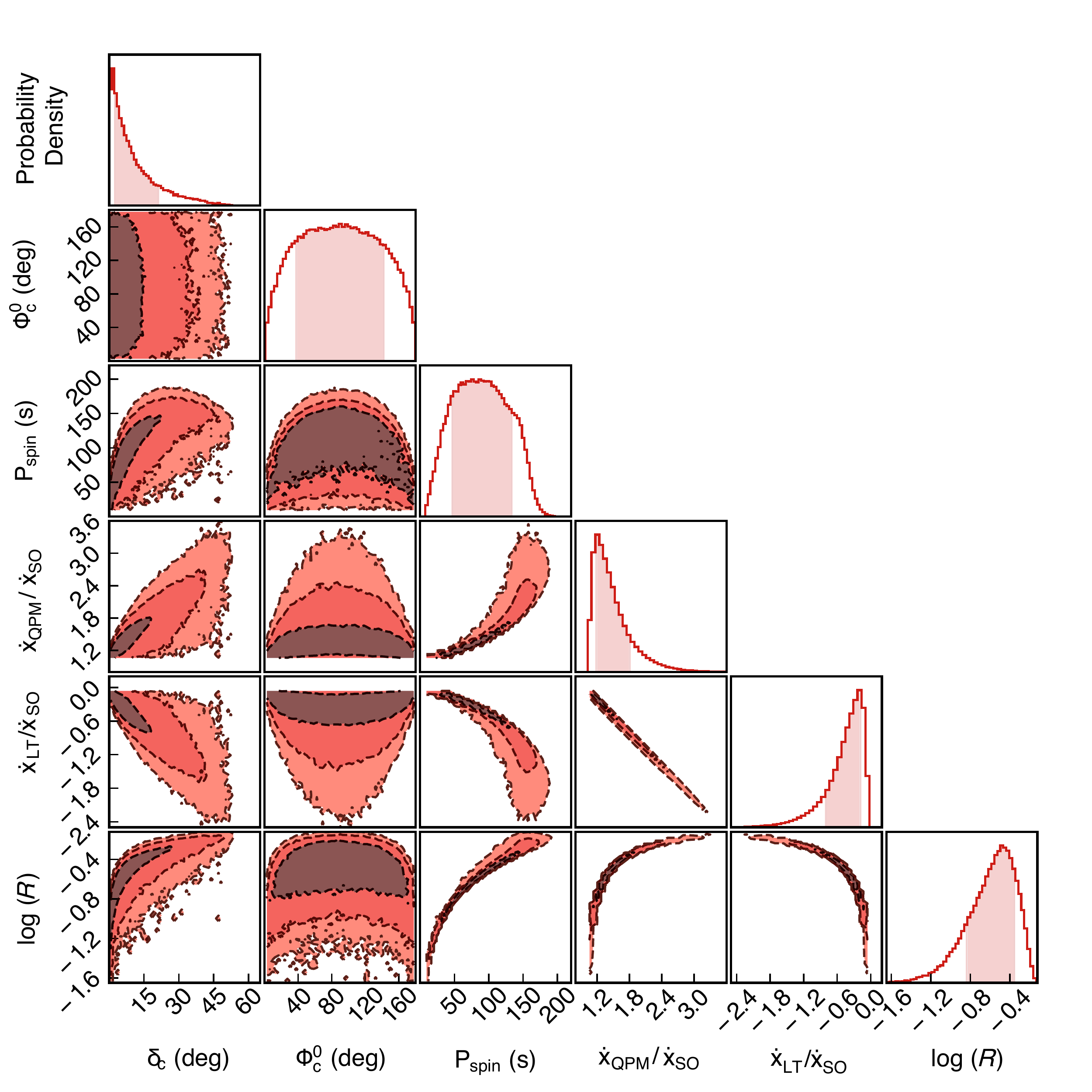} 
\end{center}
{\bf Figure~S3.} {\bf Contributions to orbital precession from WD rotation constrained by binary evolution simulations.}  Same as Figure~S2, but now with a constrained prior on $\delta_{\rm c}$ given by the binary evolution simulations (see text).

\clearpage

\begin{center}
  \includegraphics[scale=0.55,trim=20 4 4 4,clip]{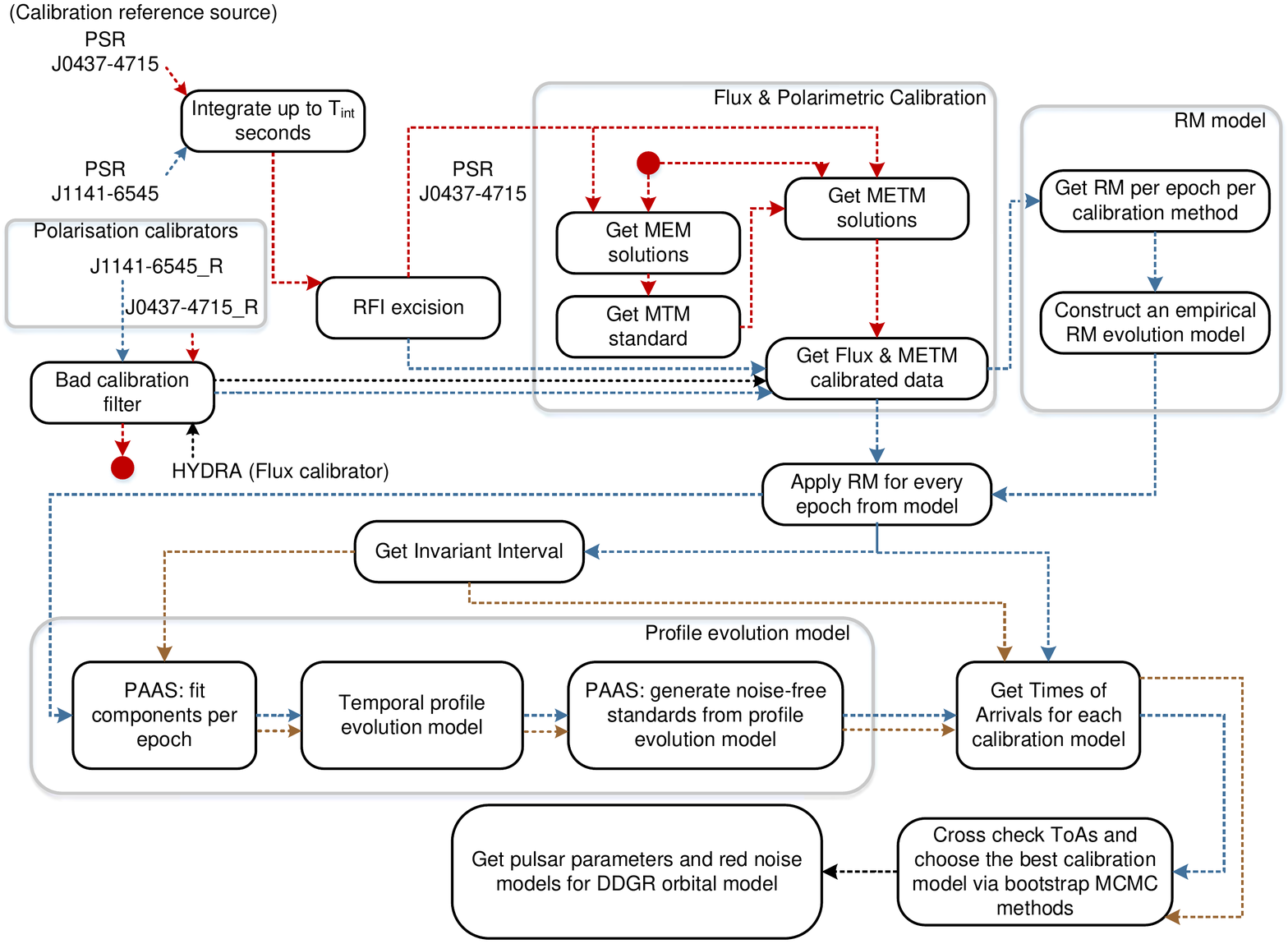} 
\end{center}
{\bf Figure~S4.} {\bf Reduction process for Parkes data.} A block diagram of the data reduction process for the Parkes 20cm multi-beam receiver data. 
\clearpage

\begin{sidewaysfigure}[ht]
  \includegraphics[scale=0.6,trim=4 4 4 4,clip]{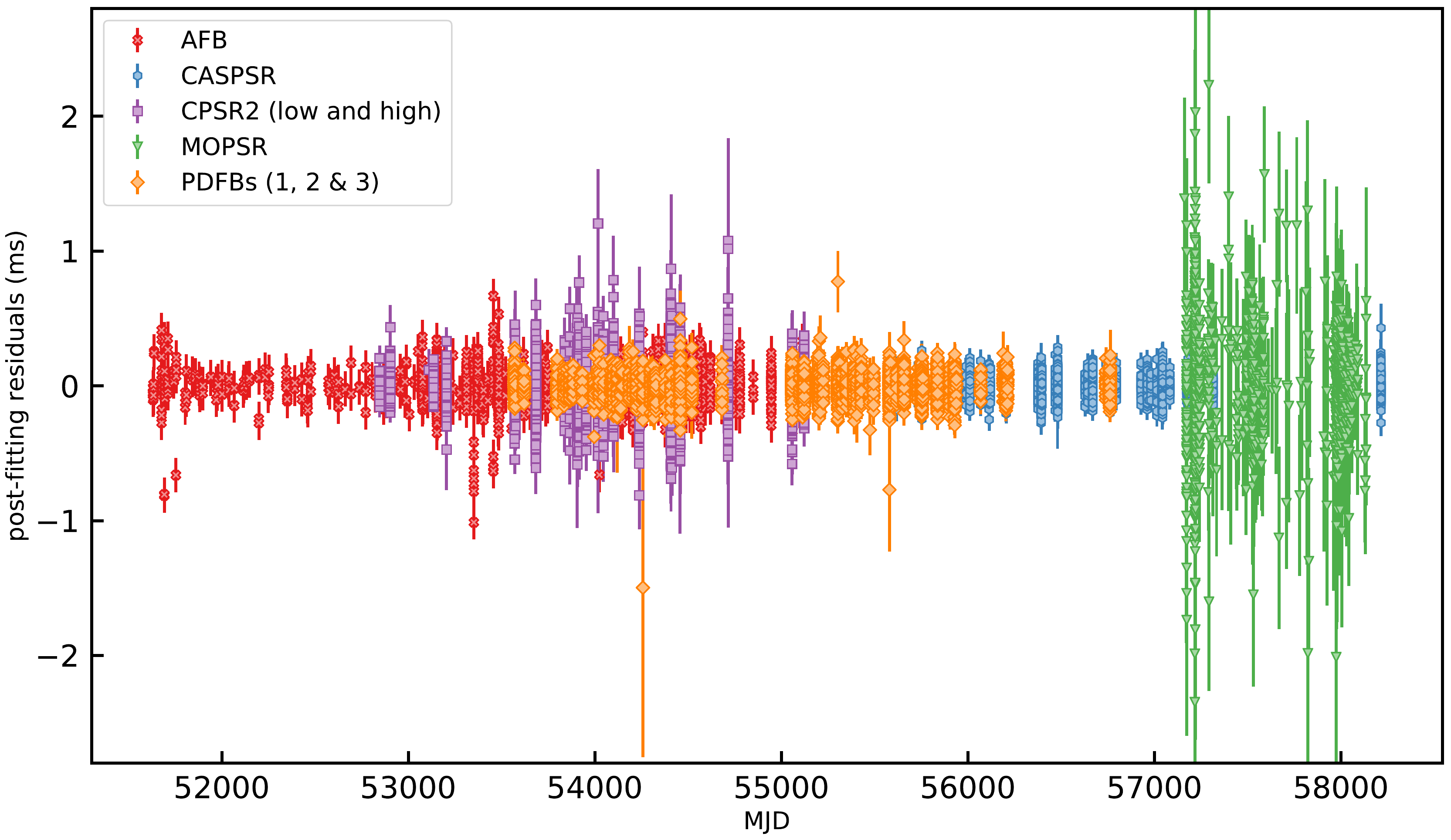} 
    \caption*{{\bf Figure~S5.} {\bf Timing residuals.} A plot of the residual errors in the arrival times of $\sim22000$ ToAs recorded over the course of our dataset. The different colors and symbols represent different backends (see text for details). The residual timing precision of the pulsar is $\sim95.6~\mu s$}
\end{sidewaysfigure}

\clearpage

\begin{center}
  \includegraphics[scale=0.39,trim= 4 4 4 4 ,clip]{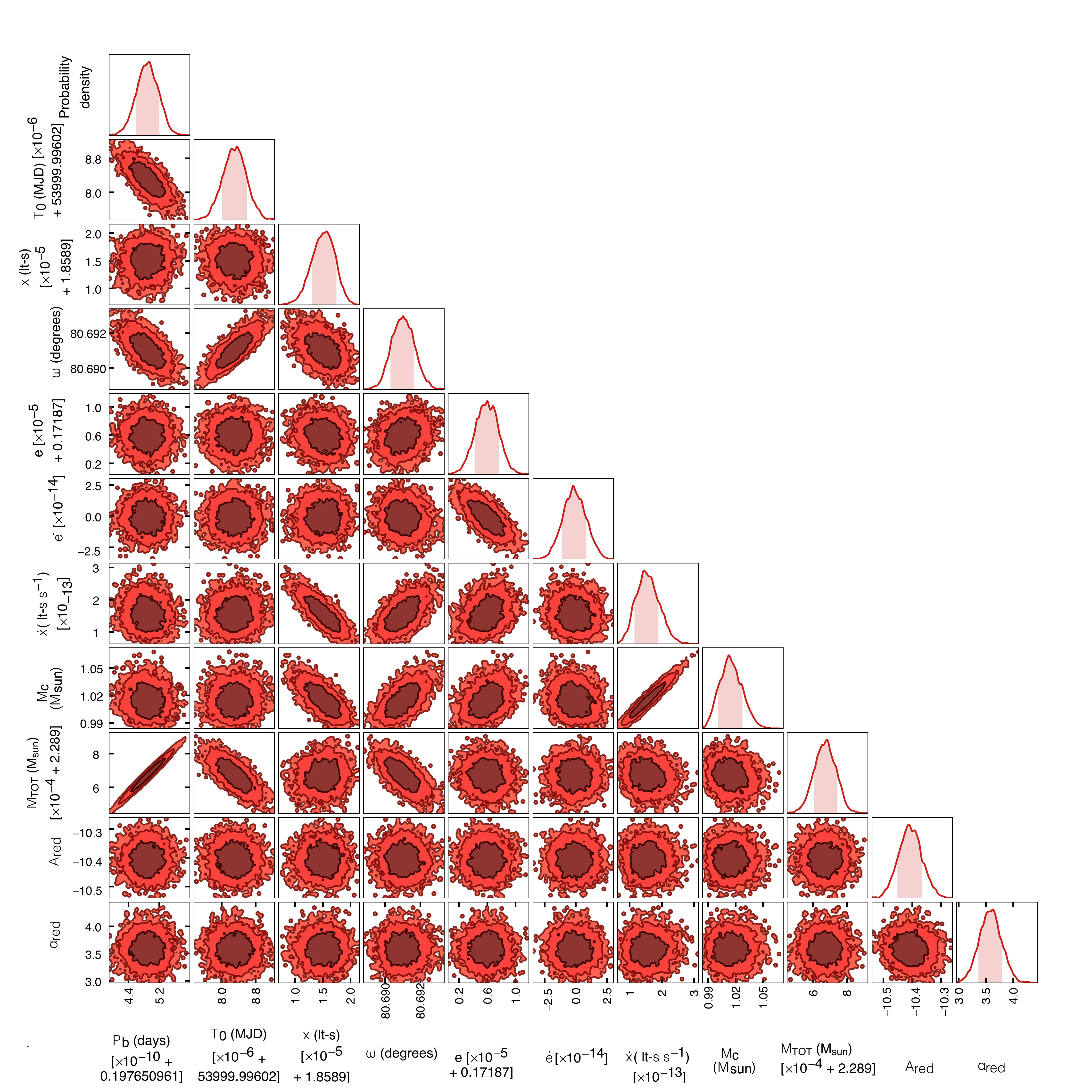} 
\end{center}
{\bf Figure~S6.} {\bf Correlations between orbital parameters.} An overview of the posterior distributions of the pulsar's DDGR orbital parameters and the red-noise model, showing the correlation between different parameters.
\clearpage 
\hspace*{-1cm}                                                           
\includegraphics[scale=0.73,trim=4 110 4 130,clip]{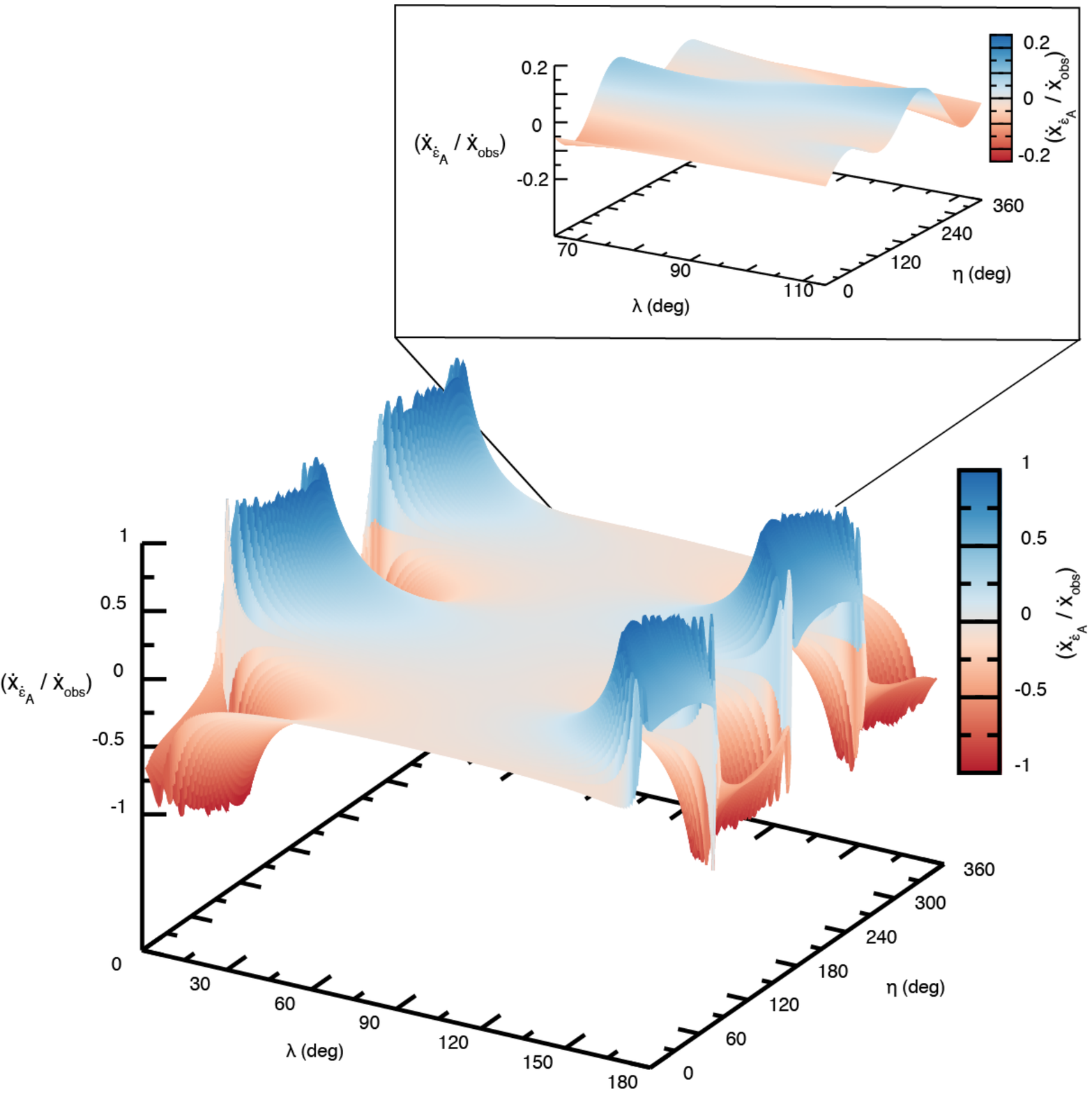} 

\noindent
{\bf Figure~S7.} {\bf Contribution to $\dot{x}_{\rm obs}$ due to changing aberration.} The fractional contribution to $\dot{x}_{\rm obs}$ from the rate of change of aberration as a function of the initial polar angles ($\lambda_{\rm p}$,$\eta_{\rm p}$) of the pulsar. As $\lambda_{\rm p} \to 0 \degree {\rm (or}\,180 \degree)$, $ \dot{x}_{\dot{\epsilon}_{\rm A}} / \dot{x}_{\rm obs} \to \pm \infty$ depending on $\eta_{\rm p}$. The vertical range is limited to $-1 <  \dot{x}_{\dot{\epsilon}_{\rm A}} / \dot{x}_{\rm obs} < 1$ for clarity. The inset shows the range of $\lambda_{\rm p}$ estimated from analysis of the temporal evolution of the observed pulse profile \cite{VenkatramanKrishnanEtAl2018}. With this constraint, it can be seen that the maximum contribution from the rate of change of aberration is only $\sim 20\%$ of $\dot{x}_{\rm obs}$ .
\clearpage

\begin{center}
  \includegraphics[scale=0.4,trim=4 4 4 4,clip]{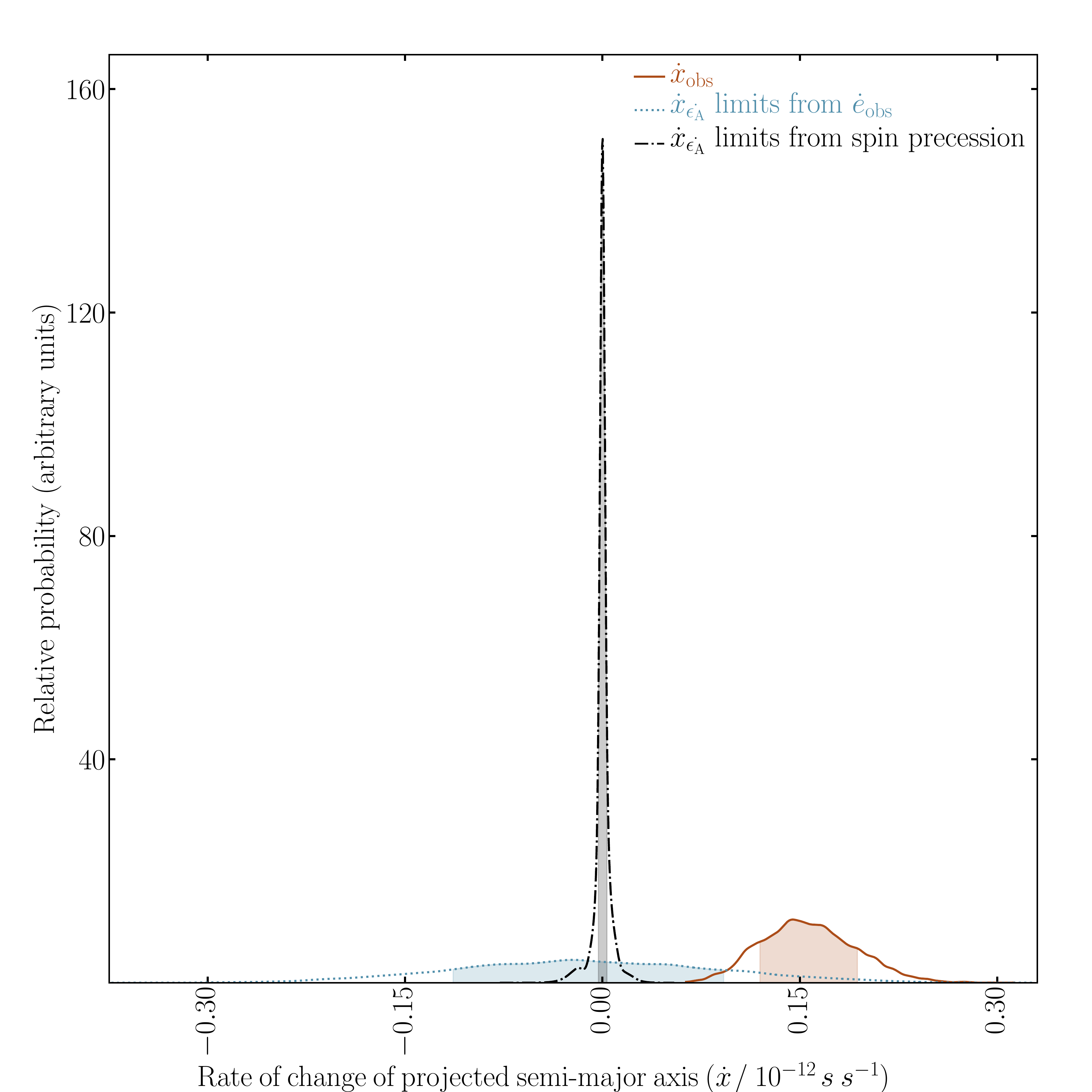} 
\end{center}
{\bf Figure~S8.} {\bf Limits on orbital precession due to changing aberration.} Limits on the contributions to the observed rate of change of the projected semi-major axis ($\dot{x}_{\mathrm{obs}}$) due to the rate of change of aberration of the pulse profile (${\dot{x}}_{\dot{\epsilon}_{\rm A}}$).  The limit in light blue comes from our non-detection of a change in eccentricity ($\dot{e}$). The more stringent limit in gray, comes from modelling the pulsar's profile evolution due to relativistic spin precession \cite{VenkatramanKrishnanEtAl2018}. The observed value,$\dot{x}_{\mathrm{obs}}$, is plotted in red. The shaded regions denote the 68\% confidence intervals. 
\clearpage

\begin{center}
  \includegraphics[width=0.95\columnwidth,angle=0]{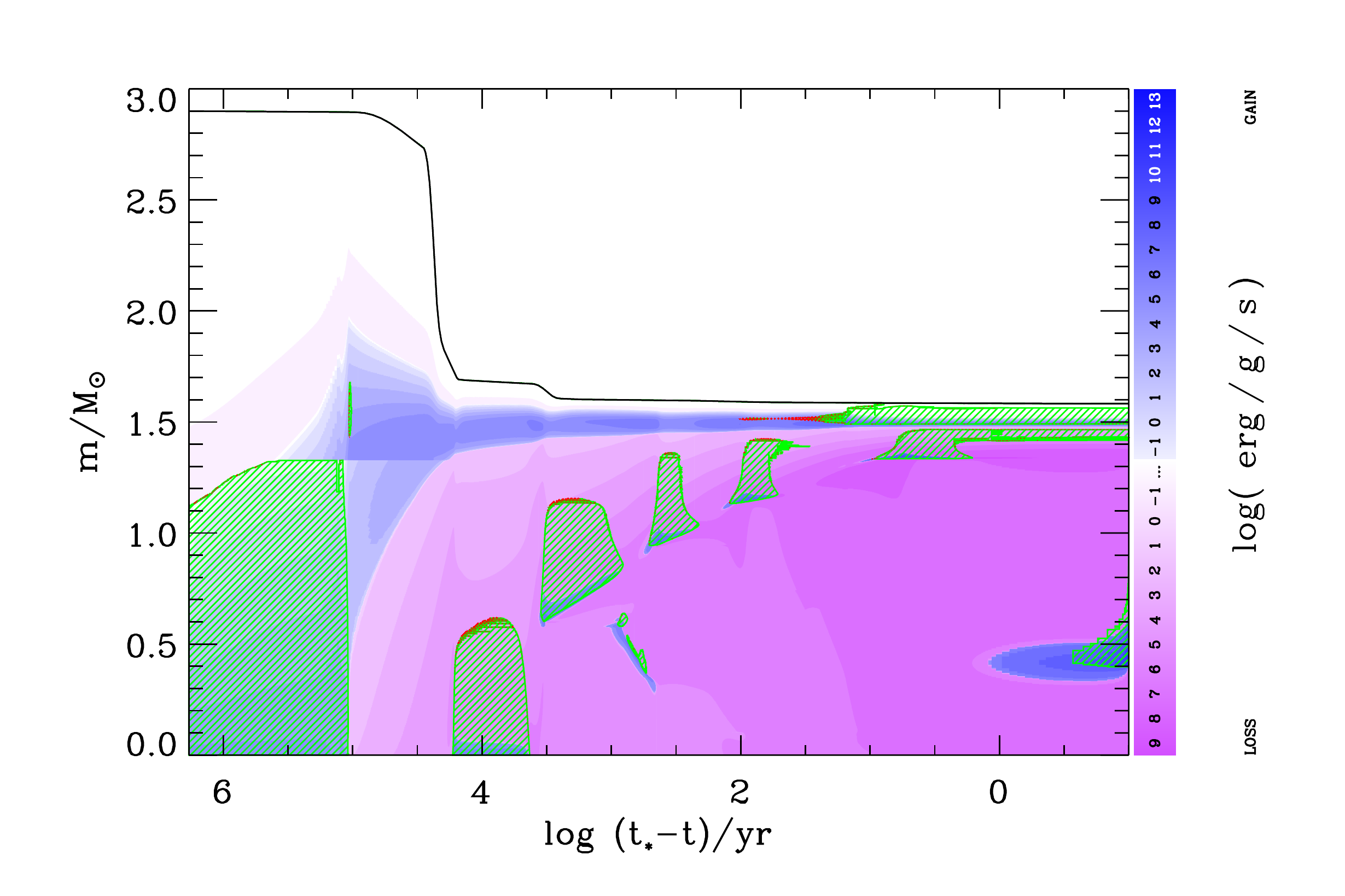}    
\end{center}
{\bf Figure~S9.} {\bf Kippenhahn diagram of a $2.9\,M_{\odot}$ helium star undergoing Case~BB RLO to a WD accretor.}
Cross-sections of the helium star are shown in mass-coordinates $(m/M_\odot)$ from the centre to the surface of the star, as a function of stellar age. The value $(t_{*}-t)/{\rm yr}$ is the remaining time of our calculations, spanning a total time of $t_*\simeq 1.8~{\rm Myr}$.
The green hatched areas denote zones with convection.
The intensity of the blue/purple regions indicates the net energy-production rate.
We expect the star to undergo an iron-core collapse about 10~yr after the off-center oxygen ignition 
(at $m/M_{\odot}\simeq 0.4$, when $\log (t_*-t)=0.0$). 

\clearpage

\begin{center}
  \includegraphics[width=0.95\columnwidth,angle=0]{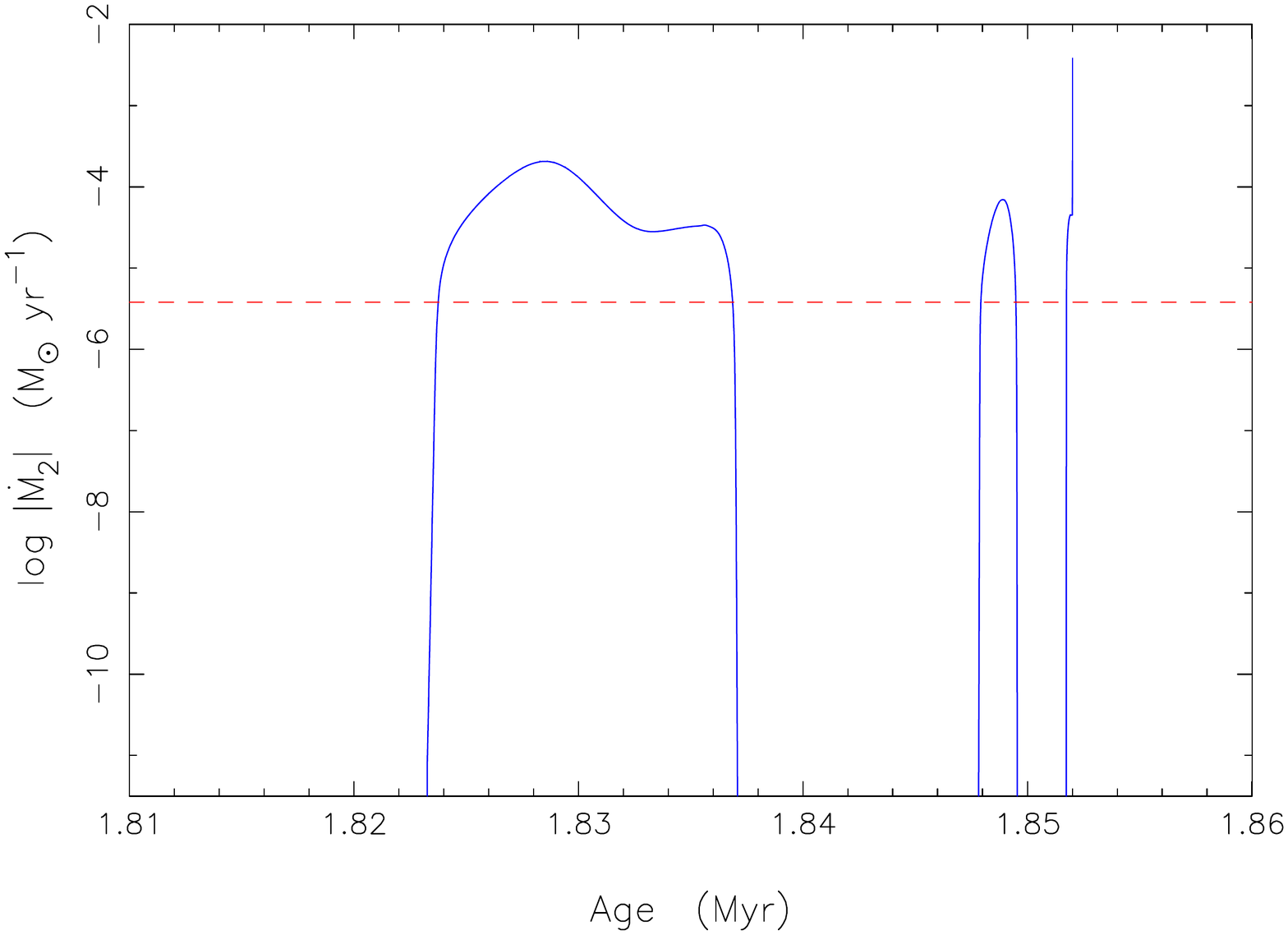}
\end{center}
{\bf Figure~S10.} {\bf Mass-transfer rate ($|\dot{M}_2|$) as a function of stellar age for the helium-star evolution plotted in Figure~S9.} During RLO, the mass-transfer rate exceeds the Eddington rate, by more than an order of magnitude, marked by the horizontal red dashed line at $|\dot{M}_2| = \dot{M}_{\rm Edd}=4\times 10^{-6}$ $M_{\odot} \rm yr^{-1}$ for the WD accretor.

\clearpage

\begin{center}
  \includegraphics[width=0.80\columnwidth,angle=0]{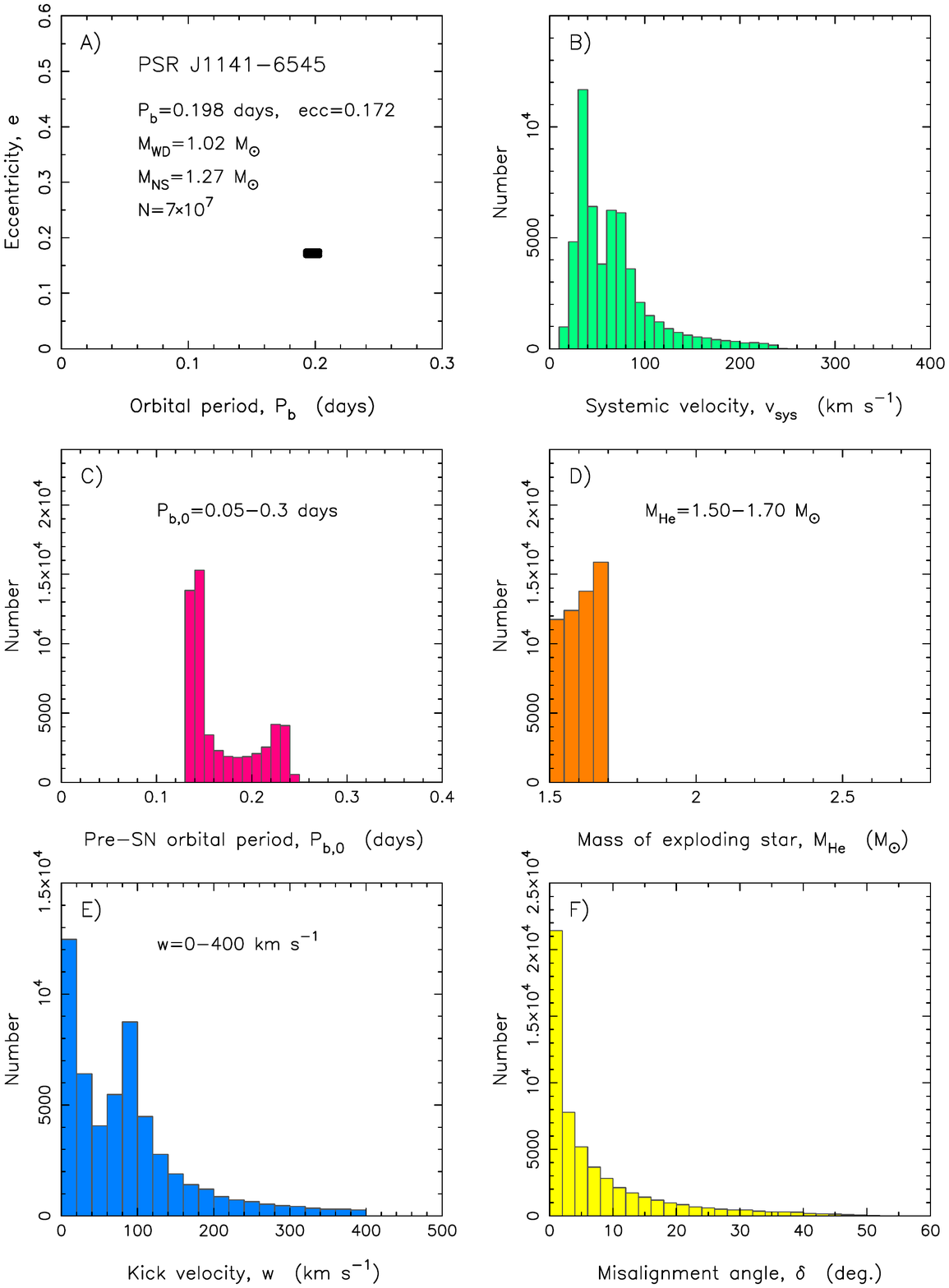}
\end{center}
{\bf Figure~S11.} {\bf Supernova explosion simulations.} Properties and constraints on the formation of the PSR J1141$-$6545 system based on Monte Carlo simulations of 70~million SN explosions following the method of \cite{Tauris+2017}. The six panels display distributions of (A--F): post-SN orbital period and eccentricity, post-SN 3D systemic velocity, pre-SN orbital period, pre-SN mass of exploding helium star, magnitude of SN kick velocity, and misalignment angle of the WD.

\clearpage

\begin{center}
\begin{table}
    \centering
    \caption*{{\bf Table S1.} {\bf Prior distributions for SN simulations}.To obtain the initial binary parameters that will produce systems that resemble PSR~J1141$-$6545, the following SN prior distributions of parameters were used.}
    \label{tab:model_posteriors}
    \begin{tabular}{ccc}
        \hline
        \hline
        \rule{0pt}{4ex}    
    Model parameter & Type of prior & Bounds \\
    \hline
    $P_{\rm b,0}$    & Uniform & by iteration - see text \\
    $M_{\rm He}$    & Uniform & [1.5,~1.7] $M_{\odot}$ \\
    $w_{\rm kick}$ & Uniform & [0,\,400] km\,s$^{-1}$ \\
    $\theta$ & $\sin \theta$ & [0,~$\pi$]\\
    $\phi$ & Uniform & [0,~$2\pi$]\\
    \hline
    \end{tabular}
\end{table}
\end{center}

\bibliographystyle{Science}

\end{document}